\documentclass[fleqn,usenatbib]{mnras}

\usepackage{newtxtext,newtxmath}

\usepackage[T1]{fontenc}

\DeclareRobustCommand{\VAN}[3]{#2}
\let\VANthebibliography\thebibliography
\def\thebibliography{\DeclareRobustCommand{\VAN}[3]{##3}\VANthebibliography}

\usepackage{graphicx}	
\usepackage{amsmath}	
\usepackage{soul}

\newcommand{\NII}{{[N\,{\sc ii}]}}

\newcommand{\NIIl}{{[N\,{\sc ii}]\,$\lambda$}}
\newcommand{\SII}{{[S\,{\sc ii}]}}

\newcommand{\SIIl}{{[S\,{\sc ii}]\,$\lambda$}}

\newcommand{\OIII}{{[O\,{\sc iii}]}}

\newcommand{\OIIIl}{{[O\,{\sc iii}]\,$\lambda$}}
\newcommand{\OII}{{[O\,{\sc ii}]}}

\newcommand{\OI}{{[O\,{\sc i}]}}

\newcommand{\objA}{GS-NDG-9422}
\newcommand{\objB}{A2744-NDG-ZD4}

\title[Nebular dominated galaxies]{Nebular dominated galaxies: insights into the stellar initial mass function at high redshift}

\author[A. J. Cameron et al.]{Alex J. Cameron,$^{1}$\thanks{E-mail: alex.cameron@physics.ox.ac.uk}
Harley Katz,$^{1,2}$
Callum Witten,$^{3,4}$
Aayush Saxena,$^{1}$
Nicolas Laporte,$^{5}$
\newauthor
and Andrew J. Bunker,$^{1}$
\\
$^{1}$Department of Physics, University of Oxford, Denys Wilkinson Building, Keble Road, Oxford, OX1 3RH, UK\\
$^{2}$Department of Astronomy \& Astrophysics, University of Chicago, 5640 S Ellis Avenue, Chicago, IL 60637, USA\\
$^{3}$Institute of Astronomy, University of Cambridge, Madingley Road, Cambridge, CB3 0HA, UK\\
$^{4}$Kavli Institute for Cosmology, University of Cambridge, Madingley Road, Cambridge, CB3 0HA, UK\\
$^{5}$Aix-Marseille Universit\'e, CNRS, CNES, LAM (Laboratoire d’Astrophysique de Marseille), UMR 7326, 13388 Marseille, France\\
}

\date{Accepted XXX. Received YYY; in original form ZZZ}

\pubyear{2024}

\begin{document}
\label{firstpage}
\pagerange{\pageref{firstpage}--\pageref{lastpage}}
\maketitle

\begin{abstract}

We identify a low-metallicity ($12+\log({\rm O}/{\rm H})=7.59$) Ly$\alpha$-emitting galaxy at $z=5.943$ with evidence of a strong Balmer jump, arising from nebular continuum. While Balmer jumps are sometimes observed in low-redshift star-forming galaxies, this galaxy also exhibits a steep turnover in the UV continuum. Such turnovers are typically attributed to absorption by a damped Ly$\alpha$ system (DLA); however, the shape of the turnover and the high observed Ly$\alpha$ escape fraction ($f_{\rm esc,Ly\alpha}~\sim27\%$) is also consistent with strong nebular two-photon continuum emission. Modelling the UV turnover with a DLA requires extreme column densities ($N_{\rm HI}>10^{23}$~cm$^{-2}$), and simultaneously explaining the high $f_{\rm esc,Ly\alpha}$ requires a fine-tuned geometry. In contrast, modelling the spectrum as primarily nebular provides a good fit to both the continuum and emission lines, motivating scenarios in which (a) we are observing only nebular emission or (b) the ionizing source is powering extreme nebular emission that outshines the stellar emission. 
The nebular-only scenario could arise if the ionising source has `turned off' more recently than the recombination timescale ($\sim$1,000~yr), hence we may be catching the object at a very specific time.
Alternatively, hot stars with $T_{\rm eff}\gtrsim10^5$~K (e.g. Wolf-Rayet or low-metallicity massive stars) produce enough ionizing photons such that the two-photon emission becomes visible. While several stellar SEDs from the literature fit the observed spectrum well, the hot-star scenario requires that the number of $\gtrsim50~{\rm M}_\odot$ stars relative to $\sim5-50~{\rm M}_\odot$ stars is significantly higher than predicted by typical stellar initial mass functions (IMFs). The identification of more galaxies with similar spectra may provide evidence for a top-heavy IMF at high redshift.

\end{abstract}




\begin{figure*}
    \centering
    \includegraphics[width=\textwidth]{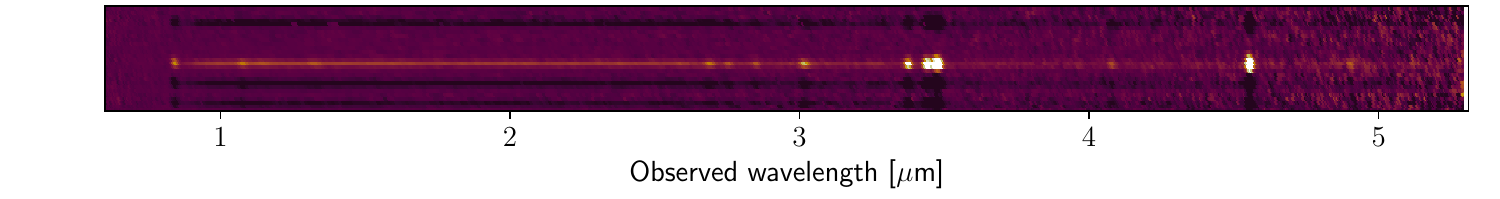}
    \includegraphics[width=\textwidth]{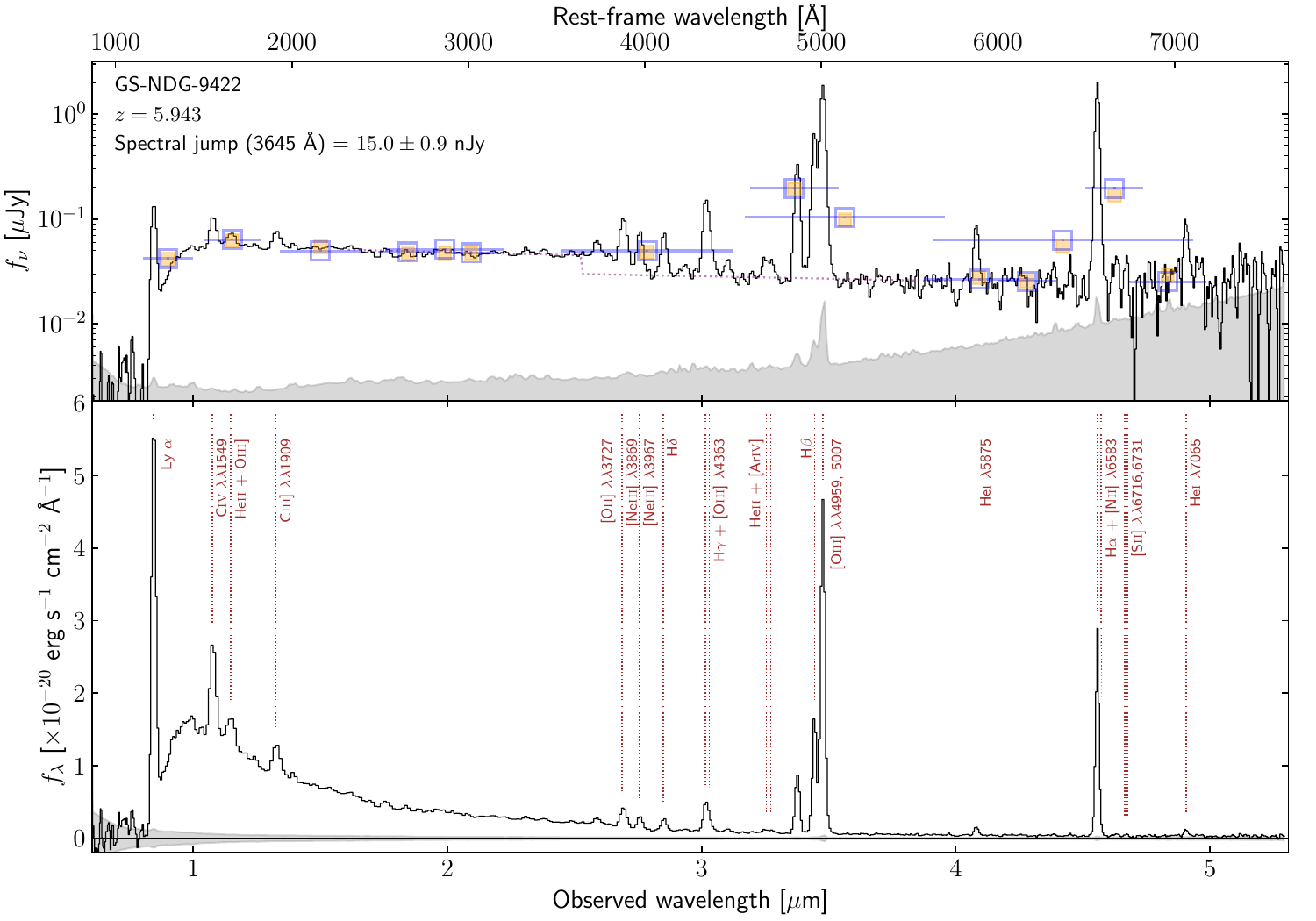}
    \includegraphics[width=\textwidth]{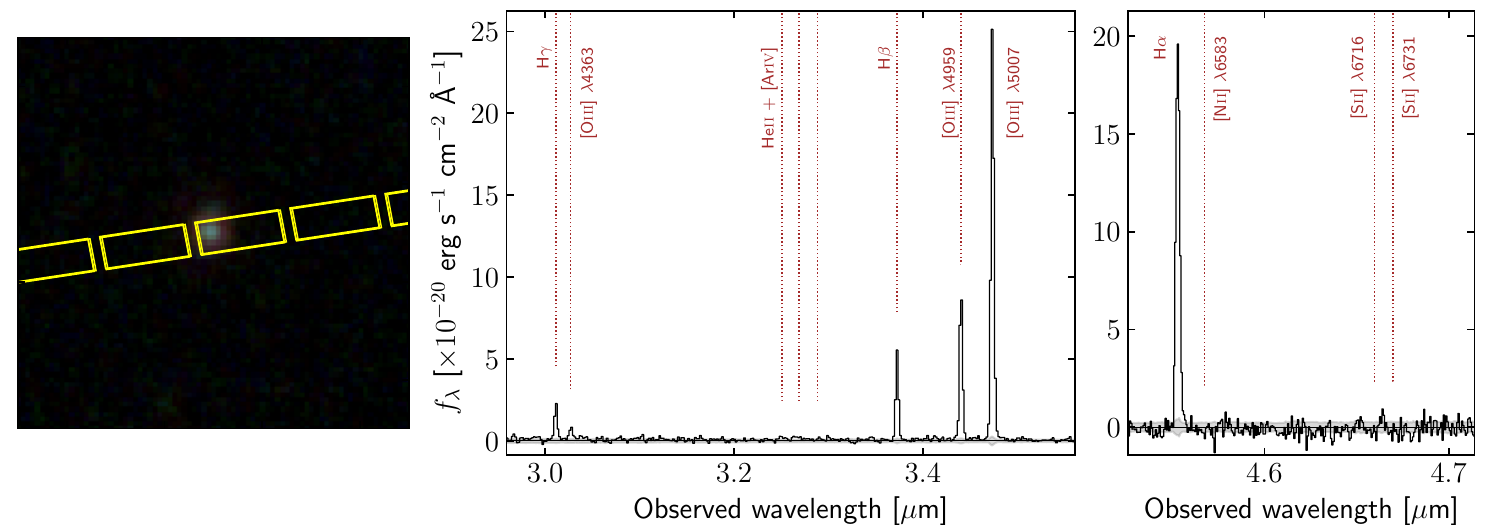}
    \caption{
    {\it Top:} 2D Prism/CLEAR spectrum of \objA. {\it Middle:} 1D Prism/CLEAR spectrum of \objA\ shown in $f_\nu$ (upper middle) and $f_\lambda$ (lower middle). Blue squares show 0.15'' radius aperture photometry from \emph{JWST}/NIRCam medium- and broad-band imaging, while orange squares show the predicted photometry obtained by convolving the Prism spectrum with the NIRCam filter transmission profiles.
    {\it Bottom left:} Three-colour image ($F090W$, $F200W$, $F444W$) of \objA\ showing the positioning of the three NIRSpec micro-shutters across the three nod positions. {\it Bottom centre:} Zoom-in of the region surrounding \OIIIl5007 in the G395M spectrum. {\it Bottom right:} Zoom-in of the region surrounding H$\alpha$ in the G395H spectrum. 
    }
    \label{fig:spectrum}
\end{figure*}

\section{Introduction} \label{sec:intro}

The unprecedented sensitivity of \emph{JWST} in the near-infrared has revolutionised our ability to study the rest-frame ultraviolet to optical spectra of high-redshift galaxies.
\emph{JWST} spectroscopy has already unveiled large samples of emission line galaxies at $z\gtrsim5$, providing new insights in to the conditions of the interstellar media (ISM) of these galaxies \citep{Matthee2023, Cameron2023_ratios, Mascia2023, Sanders2023, Boyett2024, RobertsBorsani2024}.
Furthermore, the sensitivity of \emph{JWST}/NIRSpec has even enabled high fidelity spectroscopic continuum measurements, which have provided insights into high-redshift neutral gas \citep{Heintz2023_DLA, Umeda2023}, assembly of massive galaxies in the early Universe \citep{Carnall2023, Glazebrook2023}, and have also provided the many of the highest spectroscopic redshift confirmations to date \citep{CurtisLake2023, ArrabalHaro2023}.
Spectral energy distribution (SED) fitting of many photometric data sets has indicated a need for strong nebular emission, including predictions for strong Balmer jumps \citep{Endsley2023, Topping2023}.
The strong nebular contribution implied by these fits motivates more detailed studies examining the contribution of the nebular continuum to the integrated spectra of high-redshift galaxies.


The nebular continuum is comprised of three main components: (1) free-bound emission, (2) free-free emission, and (3) two-photon emission.
Free-bound emission is the component that is most commonly observed to make a significant contribution to the integrated optical spectra of galaxies, presenting in the form of the `Balmer jump', a discontinuity in the spectrum at $\lambda_{\rm rest}=3645$~\AA{}, corresponding to the Balmer limit. This feature arises due to electron recombinations with ionized hydrogen to the first excited state.
Balmer jumps only appear in the spectra of galaxies that contain young stellar populations with high ionizing photon production efficiencies ($\xi_{\text ion}$).
They have been detected in some highly star-forming galaxies at low-redshift \citep{Peimbert1969, Guseva2006, Guseva2007}.
Many numerical simulations predict Balmer jumps to be common at high redshift \citep[e.g.][]{Katz2023_SPHINX_DR, Wilkins2023}, in line with the early SED fitting results outlined above. 
In contrast, predictions based on typical stellar population models indicate that the other nebular continuum components, free-free and two-photon emission, are generally expected to be subdominant compared to the stellar and free-bound contributions in the integrated spectra of galaxies.

The predicted subdominance of two-photon emission arises because the $\xi_{\rm ion}$ values of typical low-metallicity stellar populations are not high enough to drive nebular continuum that overcomes the steep UV stellar continuum slopes \citep{Leitherer1999}. 
In contrast, very hot stars, with blackbody temperatures of $\sim100,000$~K, produce enough ionizing photons to power nebular continuum emission that outshines their stellar UV emission, and are predicted to exhibit a strong two-photon continuum bump, peaking at $\lambda_{\rm rest}\approx1430$~\AA{} \citep{Schaerer2002, Raiter2010, Zackrisson2011, Trussler2023}.
Observations of the Lynx arc, a strongly-lensed extreme emission system at $z=3.357$, have suggested that the UV continuum of this system has a strong contribution from two-photon continuum, leading to the suggestion of the presence of hot stars with $T_{\rm eff}\sim80,000$ K \citep{Fosbury2003}.
But this has so far remained an isolated candidate for this type of system, with other examples of two-photon continuum being the domain of unusual nebular-only systems \citep[e.g.][]{Lintott2009}. 
Nonetheless, the identification of these objects, should they exist, offers powerful insight into the properties of systems with extreme ionising spectra.

In this paper, we identify one object, JADES-GS+53.12175-27.79763 (\objA\ hereafter), at $z=5.943$, that exhibits high equivalent-width emission lines as well as a strong spectral discontinuity near $\lambda_{\rm rest}=3645$~\AA{} that we interpret as a Balmer jump. The UV continuum exhibits a strong turnover which could be indicative of either a strong damped Lyman-$\alpha$ absorption (DLA) system \citep[e.g.][]{Heintz2023_DLA}, or the two-photon continuum introduced above. We present a detailed analysis of this system, outlining a number of scenarios that can potentially explain the physical origin of this intriguing spectrum. 

The paper is structured as follows: Section~\ref{sec:data} outlines the details of data used in this work. In Section~\ref{sec:physical_conditions} we present emission-line-based analysis of the nebular conditions. We perform fitting to the continuum shape in Section~\ref{sec:continuum}, comparing the results of DLA and nebular models. Section~\ref{sec:spectral_modelling} then considers the nebular case in more detail, exploring the necessary properties of the ionising source. Section~\ref{sec:discussion} outlines the broader implications of the scenarios presented. We then summarise our findings in Section~\ref{sec:conclusion}.
Throughout this paper we adopt the cosmology of \citet{Planck2016} with $H_0=67.31$~km~s$^{-1}$~Mpc$^{-1}$ and $\Omega_{\rm m}=0.315$, which gives a luminosity distance to \objA\ of $D_l=58.5$~Gpc.

\section{Data} \label{sec:data}

\emph{JWST}/NIRSpec spectroscopy of \objA\ was taken as part of the JADES survey (PID: 1210, PI: Luetzendorf) in five spectral modes, receiving 28 hours integration in Prism/CLEAR and 7 hours integration in each of  G140M/F070LP, G235M/F170LP, G395M/F290LP and G395H/F290LP. We use the reduced spectra released as part of the JADES Public Data Release \citep{Eisenstein2023, Bunker2023_DR}. 
\objA\ falls within the \emph{JWST}/NIRCam footprint of JADES (PID: 1880, PI: Eisenstein) as well as the JWST Extragalactic Medium-band Survey (JEMS; PID: 1963; PI: Williams; \citealt{Williams2023_JEMS}), combining to provide imaging in 14 wide- and medium-band filters.
In Figure~\ref{fig:spectrum} we show photometry from the publicly released photometric catalogs \citep{Rieke2023} compared to the observed Prism/CLEAR spectrum.
Blue squares in the second-top panel show aperture photometry measured within a 0.15'' radius. Orange squares show the predicted values by convolving the observed NIRSpec spectrum with the NIRCam filter transmission profiles. We find there is good agreement across the full spectral range, suggesting the flux calibration of the Prism/CLEAR spectrum is robust. The exception to this is the two filters covering H$\alpha$ emission, $F444W$ and $F460M$. In each case, the flux from imaging is 14~\% and 17~\% higher, respectively, suggesting the flux of H$\alpha$ may be marginally underestimated in the Prism/CLEAR flux calibration. 
We note that the $F460M$ flux changes by $<2$~\% when measured within apertures with radii of 0.10'' or 0.25'', suggesting that this offset is not driven by spatial variations in H$\alpha$.

\subsection{Emission line fitting}
\label{sub:line_fitting}

Where possible, emission lines were fit with a single component Gaussian profile with the continuum modelled as a 1D spline. In cases where lines are sufficiently blended, we fit the whole complex with one component. In some cases, partially blended lines are fit simultaneously with neighbouring lines and fluxes reported separately. These are marked in Table~\ref{tab:line_fluxes}.
We fit all identifiable lines and report upper limits for notable undetected lines.

Line fluxes from higher resolution grating spectra of \objA\ were measured independently. 
Emission line widths are only spectrally resolved in the high-resolution G395H grating. These show no evidence of a broad component and are well fit with a single component with velocity dispersions $<200$ km s$^{-1}$ (Table~\ref{tab:line_EWs}).

Fluxes derived from different observations are mildly discrepant in some cases. Notably, H$\beta$, [O{\sc iii}]~$\lambda$4959, [O{\sc iii}]~$\lambda$5007 and H$\alpha$ lines exhibit higher fluxes in the grating modes. This behaviour is reported in \citet{Bunker2023_DR} who suggest that the Prism flux calibration is more reliable. This conclusion is supported by the good agreement observed in Figure~\ref{fig:spectrum} between the Prism spectrum and NIRCam photometry, with the possible exception of H$\alpha$.
We note that in low-resolution data, the continuum level for some emission lines can be difficult to determine, especially for Ly$\alpha$, He~{\sc ii}~+~O~{\sc iii}] and [O~{\sc ii}]~$\lambda\lambda$3727, which introduces uncertainty into the emission line flux.
The clear detection of the continuum across almost the entire Prism coverage, allows us to derive equivalent widths directly from this spectrum (Table~\ref{tab:line_EWs}). Given the noted discrepancy on the H$\alpha$ flux between the imaging and spectroscopy, we also derive EW$_0$(H$\alpha$) directly from the imaging. Comparing the measured flux in $F460M$ ($196\pm5$ nJy), shown in Figure~\ref{fig:spectrum} to be clearly elevated due to contamination from H$\alpha$, with $F430M$ ($25\pm4$ nJy), which captures clean continuum between He~{\sc i}~$\lambda$5875 and H$\alpha$, we find the imaging implies EW$_0$(H$\alpha) = 2195\pm400$ \AA{}. 
Within the uncertainty, the measured difference between EW$_0$(H$\alpha)_{\rm Prism}$ and EW$_0$(H$\alpha)_{\rm Imaging}$ (Table~\ref{tab:line_EWs}) is consistent with the discrepancy noted above between the observed $F460M$ flux, and the predicted $F460M$ flux obtained by convolving the Prism spectrum with the NIRCam filter profile.
Throughout our analysis, we adopt the Prism fluxes where possible. However, we ensure that conclusions presented in this work are also consistent with measured grating 
flux ratios.

\begin{table*}
\caption{Emission line flux measurements for \objA\ across each observed spectrum. Fluxes are reported in units of  $\times10^{-19}$ erg s$^{-1}$ cm$^{-2}$.}
\begin{tabular}{lccccc}
\hline
 & Prism/CLEAR & G140M & G235M & G395M & G395H \\
\hline
\rm Ly-$\alpha$ & $110.1\pm2.8$ & $101.4\pm4.3$ & ... & ... & ... \\
\rm N{\sc v}~$\lambda$1239 &  & $<5.4$ & ... & ... & ... \\
\rm N{\sc v}~$\lambda$1243 &  & $<4.9$ & ... & ... & ... \\
\rm N{\sc iv}]~$\lambda$1483 &  & $<2.4$ & ... & ... & ... \\
\rm N{\sc iv}]~$\lambda$1486 &  & $<3.1$ & ... & ... & ... \\
\rm C{\sc iv}~$\lambda$1548 &  & $21.5\pm1.1^\dagger$ & ... & ... & ... \\
\rm C{\sc iv}~$\lambda$1550 &  & $17.6\pm1.1^\dagger$ & ... & ... & ... \\
\rm ~C{\sc iv}~$\lambda\lambda$1549 & $37.8\pm0.9$ &  &  &  &  \\
\rm He{\sc ii}~$\lambda$1640 &  & $5.1\pm0.9$ & ... & ... & ... \\
\rm O{\sc iii}]~$\lambda$1660 &  & $<2.5^\dagger$ & ... & ... & ... \\
\rm O{\sc iii}]~$\lambda$1666 &  & $4.9\pm1.0^\dagger$ & ... & ... & ... \\
\rm ~He{\sc ii}+O{\sc iii}] & $15.1\pm1.1$ &  &  &  &  \\
\rm N{\sc iii}]~$\lambda\lambda$1750 &  & $<1.9$ & ... & ... & ... \\
\rm C{\sc iii}]~$\lambda\lambda$1909 & $13.6\pm0.9$ & $8.0\pm1.0$ & ... & ... & ... \\
\rm [O{\sc ii}]~$\lambda\lambda$3727 & $2.2\pm0.2$ & ... & $3.0\pm0.6$ & ... & ... \\
\rm H11 & ... & ... & $<1.0$ & ... & ... \\
\rm H10 & ... & ... & $<0.8$ & ... & ... \\
\rm H9 & ... & ... & $1.2\pm0.3$ & ... & ... \\
\rm [Ne{\sc iii}]~$\lambda$3869 &  & ... & $6.2\pm0.5$ & ... & ... \\
\rm He{\sc i}~$\lambda$3889 &  & ... & $1.6\pm0.4$ & ... & ... \\
\rm ~[Ne{\sc iii}]+He{\sc i} & $7.6\pm0.4$ &  &  &  &  \\
\rm H$\delta$ & $4.3\pm0.5$ & ... & ... & ... & ... \\
\rm H$\gamma$ & $8.3\pm0.4^\dagger$ & ... & $7.5\pm0.5$ & $8.1\pm0.5$ & $5.9\pm0.6$ \\
\rm [O{\sc iii}]~$\lambda$4363 & $2.8\pm0.4^\dagger$ & ... & $2.6\pm0.7$ & $3.4\pm0.5$ & $2.8\pm0.5$ \\
\rm He{\sc i}~$\lambda$4471 & $0.8\pm0.1$ & ... & $<3.1$ & $<14.6$ & $2.1\pm0.5$ \\
\rm He{\sc ii}~$\lambda$4686 & $1.1\pm0.2^\dagger$ & ... & ... & $<1.5$ & $<2.4$ \\
\rm [Ar{\sc iv}]~$\lambda$4711$^\ddag$ & $0.8\pm0.2^\dagger$ & ... & ... & $1.5\pm0.4$ & $<1.3$ \\
\rm [Ar{\sc iv}]~$\lambda$4740 & $0.4\pm0.2^\dagger$ & ... & ... & $<0.8$ & $<1.2$ \\
\rm H$\beta$ & $17.2\pm0.3$ & ... & ... & $19.4\pm0.5$ & $19.6\pm0.5$ \\
\rm [O{\sc iii}]~$\lambda$4959 & $32.3\pm0.4$ & ... & ... & $36.3\pm0.7$ & $36.4\pm0.6$ \\
\rm [O{\sc iii}]~$\lambda$5007 & $97.1\pm0.8$ & ... & ... & $102.2\pm1.1$ & $98.7\pm1.5$ \\
\rm He{\sc i}~$\lambda$5875 & $2.2\pm0.1$ & ... & ... & $1.3\pm0.4$ & ... \\
\rm [O{\sc i}]~$\lambda$6300 & $<0.5$ & ... & ... & $<1.7$ & $<1.3$ \\
\rm H$\alpha$ & $45.5\pm0.6$ & ... & ... & ... & $52.5\pm0.9$ \\
\rm [N{\sc ii}]~$\lambda$6583 & ... & ... & ... & ... & $<1.4$ \\
\rm [S{\sc ii}]~$\lambda$6716 & $<0.5$ & ... & ... & ... & $<1.6$ \\
\rm [S{\sc ii}]~$\lambda$6731 & $<0.5$ & ... & ... & ... & $<1.7$ \\
\rm He{\sc i}~$\lambda$7065 & $1.2\pm0.1$ & ... & ... & $<2.9$ & $<2.1$ \\
\hline
\end{tabular}
\\
\raggedright
$^\dagger$ Fit simultaneously with neighbouring line and fluxes reported separately.\\
$^\ddag$ Blended with He~{\sc i} $\lambda$4713.
\label{tab:line_fluxes}
\end{table*}

\begin{table}
\caption{Rest-frame equivalent widths of prominent lines measured from the Prism/CLEAR spectrum, and line widths measured from $R\sim2700$ G395H/F290LP grating observations. We also reported the equivalent width of H$\alpha$ measured from $F460M$ and $F430M$ photometry.}
\centering
\begin{tabular}{lccc}
\hline
Line & EW$_0$ (\AA{}) & EW$_0$ (\AA{}) & FWHM (km s$^{-1}$)  \\
 & Prism/CLEAR & Imaging & G395H/F290LP  \\
\hline
Ly-$\alpha$ & $328\pm25$ &  & \\
H$\beta$ & $383\pm29$ &  & $184 \pm 5$ \\
\OIIIl5007 & $2275\pm159$ &  & $177 \pm 3$ \\
H$\alpha$ & $1784\pm340$ & $2195\pm400$ & $161 \pm 3$ \\
\hline
\end{tabular}
\label{tab:line_EWs}
\end{table}

\begin{figure*}
    \centering
    \includegraphics[width=\textwidth]{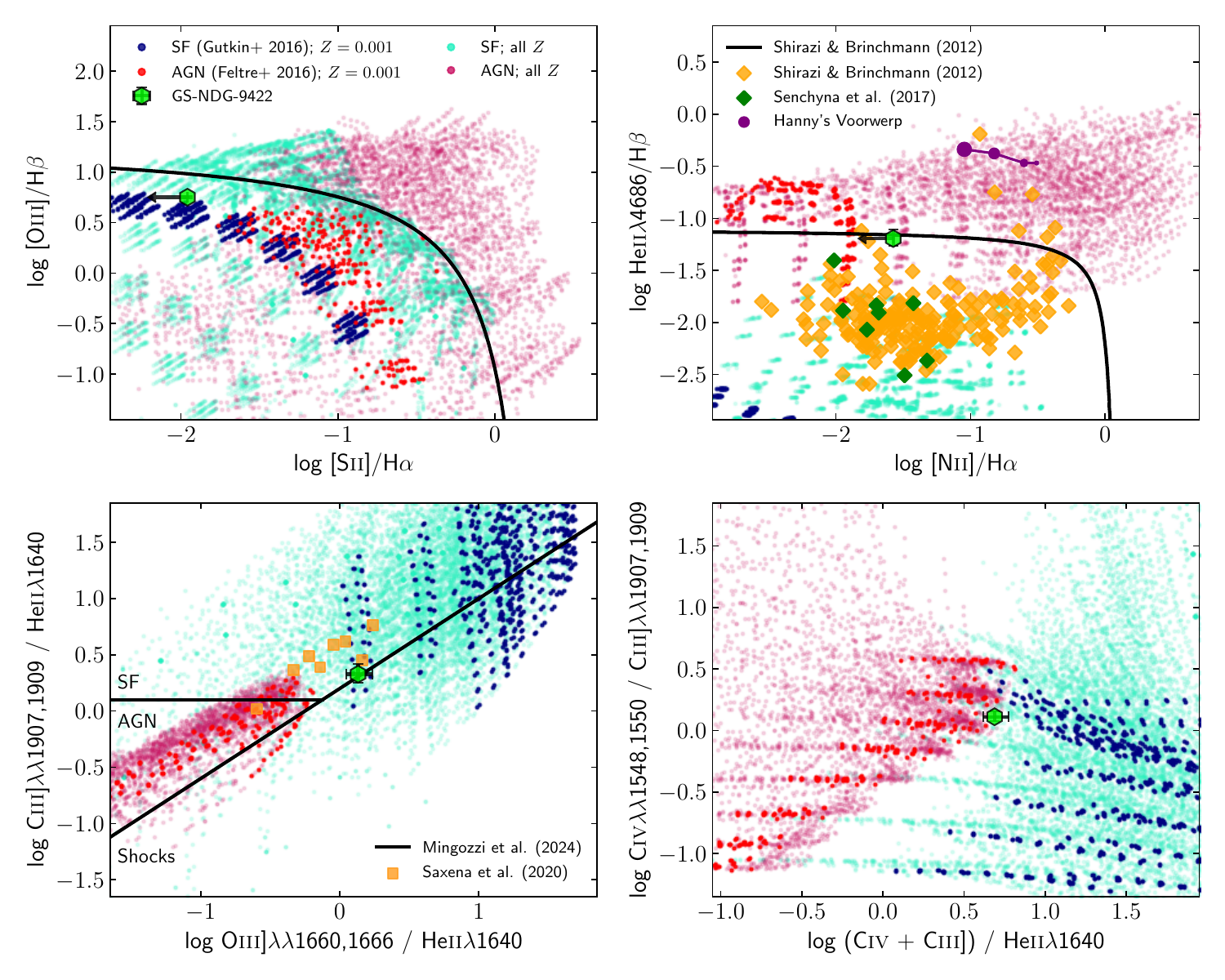}
    \caption{\objA\ plotted onto several line ratio diagnostic diagrams. Light blue points show model predictions for star-forming regions from \citet{Gutkin2016}, while pink points show AGN model predictions from \citet{Feltre2016} across a large range of metallicities. Navy (star-forming) and red (AGN) points show the subsets of these model grids that have $Z=0.001$ ($Z\approx 0.07 Z_\odot$; $12+\log ({\rm O}/{\rm H})\approx7.54$) which is the closest grid value to that measured for \objA\ (see Section~\ref{sub:chem_abun}). {\it Top left:} The strong upper limit on \SII\ $\lambda\lambda$6716, 6731 positions \objA\ well below the theoretical maximum starburst limit from \citet{Kewley2001} (black line). {\it Top right:} The weak detection of He {\sc ii} $\lambda$4686 is also below the maximum starburst limit from \citet{ShiraziBrinchmann2012} (black line). Strong He {\sc ii}-emitting star-forming galaxies from \citet{ShiraziBrinchmann2012} and \citet{Senchyna2017} are also shown as orange and green diamonds respectively. We also show measurements from Hanny's Voorwerp, suggested to be a quasar light echo (\citealt{Lintott2009}; purple line-connected points), which we discuss in Section~\ref{sub:geometry}. Points of increasing size indicate larger offset from the quasar ranging from 13 to 31 kpc.
    {\it Bottom left:} \objA\ lies beyond the AGN region defined by \citet{Mingozzi2024} (black lines), and is more coincident with the star-forming models than AGN models. He {\sc ii}-emitting star-forming galaxies at $z\sim2.5-4$ from \citet{Saxena2020_HeII} are shown for comparison as orange squares.
    {\it Bottom right:} \objA\ lies at the tip of the parameter space covered by AGN models, while it is well within the bounds of that covered by star-forming models.
    }
    \label{fig:diagnostic_diagrams}
\end{figure*}

\begin{figure}
    \centering
    \includegraphics[width=0.49\textwidth]{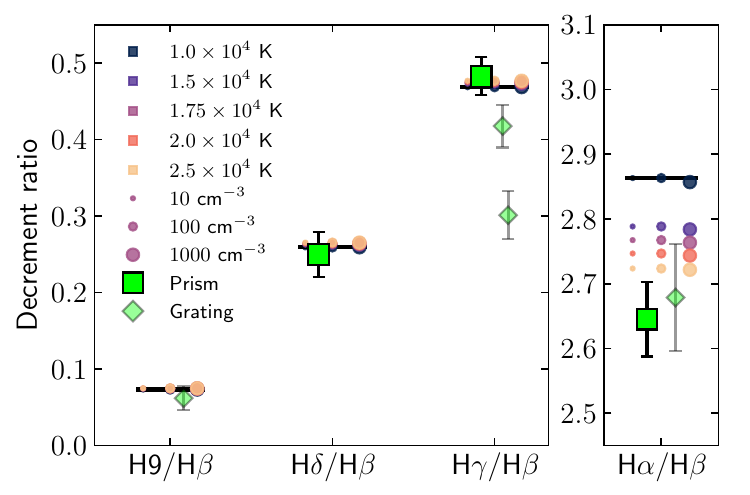}
    \caption{
    Observed Balmer decrements compared to theoretical values at different temperatures and densities. Green squares give the measured ratios from the Prism, while diamonds give ratios measured from gratings, where available. Solid black lines give the theoretical ratio for case B recombination at $T_e=10^4$ K and $n_e=100$ cm$^{-3}$. Coloured points show theoretical ratios for a range of temperatures ($10^4\leq T_e \leq 2.5 \times 10^4$ K; colour) and densities ($n_e=10$, 100, 1000 cm$^{-3}$; marker size). The measured H$\alpha$/H$\beta$ is somewhat lower than the theoretical value, but we note that H$\alpha$ may be underestimated in the prism spectrum (Table~\ref{tab:line_EWs}).}
    \label{fig:balmer_decrement}
\end{figure}

\section{Derivation of physical conditions}
\label{sec:physical_conditions}

We now explore the physical conditions of the gas in \objA\ and the basic properties of the ionising source, based on the measured emission line fluxes.
Throughout this section, we make use of {\sc pyneb} \citep{Luridiana2015_pyneb} using atomic data from {\sc chianti} \citep[version 10.0.2;][]{Dere1997_chianti, DelZanna2021_chianti}.

\subsection{Diagnostic diagrams}
\label{sub:diagnostic_diagrams}

To explore the properties of the ionising source, in Figure~\ref{fig:diagnostic_diagrams}, we look at nebular line ratio diagnostic diagrams, comparing measured line ratios from \objA\ with photoionisation model predictions for star formation and active galactic nuclei.
For star forming models, we adopt the predictions of \citet{Gutkin2016}. These use input stellar SEDs based on plane-parallel non-local thermal equilibrium models calculated with Tlusty \citep[e.g.][]{Lanz2007}. Stellar abundance patterns are assumed to follow scaled-solar, which may not be representative of $z\sim6$ stellar populations, while nebular abundance patterns allow for some variation in C/O and N/O ratios.
Star forming models with $Z=0.001$ (the grid value closest to the gas-phase oxygen abundance derived for \objA; see Section~\ref{sub:chem_abun}) are shown in dark blue, while all other star-forming models are shown in light blue.
Model predictions for active galactic nuclei from \citet{Feltre2016} are shown in red ($Z=0.001$) and pink (all other $Z$).

The non-detection of \SIIl$\lambda$6716, 6731 places \objA\ firmly below the \citet{Kewley2001} maximum-starburst limit of the classical \SII-VO87 diagram (\citealt{VO1987}; top left panel of Figure~\ref{fig:diagnostic_diagrams}) and in a region that is difficult to reconcile with emission powered by an AGN. Non-detections of \NIIl6583 and \OI\ $\lambda$6300 paint a similar picture in the \NII-BPT diagram \citep{BPT1981} and the \OI-VO87 diagram, although these are not shown.

The top right panel of Figure~\ref{fig:diagnostic_diagrams} shows that the He~{\sc ii} $\lambda$4686 / H$\beta$ ratio of \objA\ exceeds that predicted by any of the star-forming models in \citet{Gutkin2016}, especially those at $Z=0.001$. However, it is well known that, at low metallicity, star-forming galaxies are often observed with He~{\sc ii} emission exceeding that which can be powered by standard stellar population models \citep{ShiraziBrinchmann2012, Kehrig2015, Senchyna2017, Schaerer2019, Saxena2020_HeII}. He~{\sc ii}-selected star-forming galaxies from \citet{ShiraziBrinchmann2012} and \citet{Senchyna2017} are shown as orange and green diamonds respectively. We find that while \objA\ is at the upper end of the He~{\sc ii}~$\lambda$4686 / H$\beta$ distribution from these works, it does not exceed the very highest ratios, and also falls below the He~{\sc ii} maximum-starburst demarcation presented in \citet{ShiraziBrinchmann2012}.
The origin of strong He~{\sc ii} emission at low metallicity remains an unsolved problem with possible solutions including revised stellar wind properties, X-ray binaries, very hot binary stellar evolution products, rotating massive low-metallicity stars, and/or a top-heavy stellar initial mass function \citep{Kehrig2015, Kehrig2018, Schaerer2019, Senchyna2020, Olivier2022}. We return to this question in Section~\ref{sec:spectral_modelling}.
We note that \objA\ was included in a selection of narrow-line AGN in \citet{Scholtz2023} on the basis of the He~{\sc ii} $\lambda$4686 / H$\beta$ ratio. However, we consider the measured value of this line ratio to be inconclusive.

In the bottom left panel of Figure~\ref{fig:diagnostic_diagrams}, we see that the O{\sc iii}] $\lambda\lambda$1660, 1666 / He~{\sc ii} $\lambda$1640 ratio measured for \objA\ exceeds the AGN locus defined by \citet{Mingozzi2024} and has a value which cannot be reproduced by the \citet{Feltre2016} AGN models. The measured ratio is more in line with that measured in the $z\sim2-4$ He~{\sc ii}-selected star-forming galaxies from \citet{Saxena2020_HeII}. Meanwhile the (C~{\sc iv} + C~{\sc iii}]) / He~{\sc ii} ratio in the bottom right panel is only reproduced by the very tip of the AGN model parameter space, being more readily reproduced by the star-forming models.

In summary, AGN photoionisation models struggle to reproduce a number of the observed emission line ratios, especially those involving low-ionisation emission lines. In contrast, \objA\ resides within regions of line-ratio space consistent with emission powered by stars. We therefore conclude that the ionising source in \objA\ is most likely of stellar origin.

\subsection{Balmer decrements and dust extinction}
\label{sub:dust_extinction}

Balmer decrements H$\delta$/H$\beta$ and H$\gamma$/H$\beta$ from the Prism and H9/H$\beta$ from the grating are consistent with Case B values, indicating that there is no significant dust reddening in \objA\ (Figure~\ref{fig:balmer_decrement}). 
Measured H$\alpha$/H$\beta$ ratios are lower than theoretically predicted for Case B at $T=10^4$~K. Note, this is not suggestive of dust reddening, which would act in the opposite direction.
At higher temperatures, the theoretical ratio decreases, and our G395H/F290LP measurement is consistent with theoretical ratios with $T_e\gtrsim 2\times 10^4$~K, possibly indicative of a very hot nebula. 
As noted in Section~\ref{sub:line_fitting}, H$\alpha$ may be underestimated in the Prism which could lead to the slightly lower observed ratio.
H$\gamma$/H$\beta$ measured from the medium-resolution grating is marginally below the theoretical value. In isolation, this could suggest non-zero dust reddening; however, this evidence is outweighed by the other measured ratios.
The high-resolution grating returns a much lower H$\gamma$/H$\beta=0.3\pm0.03$. However, we note that the continuum is undetected in the high-resolution grating, which contributes uncertainty to the measured ratio.

Adopting a luminosity distance of $D_l=58.5$~Gpc, we derive $L_{\rm H\alpha}=1.86\times10^{42}$ erg s$^{-1}$. Assuming no dust, a metallicity of $Z=0.1 Z_\odot$, and a typical IMF, this would correspond to a star formation rate of $\sim3.2$ M$_\odot$ yr$^{-1}$ \citep{Eldridge2017}.
Under the assumption of no dust, Case B recombination, $n_e=100$ cm$^{-3}$ and $T_e=1.8\times10^4$ K, we derive a Ly$\alpha$ escape fraction of $f_{\rm esc,Ly\alpha}=0.29\pm0.01$ from the Ly$\alpha$/H$\alpha$ ratio, or $f_{\rm esc,Ly\alpha}=0.27\pm0.01$, measured from the Ly$\alpha$/H$\beta$ ratio, the latter of which may be more reliable owing to the H$\alpha$ flux uncertainty discussed above.

\subsection{Electron temperature}
The temperature-sensitive [O~{\sc iii}] $\lambda$4363/$\lambda$5007 ratio can be measured from each of the Prism, G395M and G395H observations, yielding three consistent, independent temperature measurements ($T_e = 1.83\pm0.15$, $1.99\pm0.18$, and $1.81\pm0.18\times 10^4$ K, respectively).
The temperature derived from the medium-resolution O~{\sc iii}] $\lambda$1666 / [O~{\sc iii}] $\lambda$5007 ratio is somewhat lower ($T_e=1.70^{+0.05}_{-0.06}\times10^4$~K). However, the He~{\sc ii}+O~{\sc iii}] flux measured from the medium-resolution G140M grating is significantly lower than that of the Prism.
Instead, the measured temperature from above ($T_e=1.83\times10^4$~K) implies $\lambda\lambda$1660,1666/$\lambda5007=0.08$, which gives $f_{\lambda\lambda1660,1666}=8.2\pm0.1\times10^{-19}$~erg~s$^{-1}$~cm$^{-2}$ based on the [O~{\sc iii}]~$\lambda$5007 Prism flux, suggesting that O~{\sc iii}] contributes $\sim$55~\% of the measured Prism He~{\sc ii}+O~{\sc iii}] blend. This is consistent with the O~{\sc iii}] / He~{\sc ii} ratio measured in G140M.
The implied He~{\sc ii} $\lambda$1640 flux ($6.9\pm1.1\times10^{-19}$~erg~s$^{-1}$~cm$^{-2}$) 
gives a He~{\sc ii}~$\lambda$1640/$\lambda$4686 ratio of $6.3\pm1.5$ which is consistent with the theoretical value of 7.19 assuming Case B at $T_e = 1.8\times10^4$ K and $n_e=100$ cm$^{-3}$.
Note that this further supports the conclusion of a lack of dust in this system since the He~{\sc ii}~$\lambda$1640 would be subject to extremely high extinction, relative to the $\lambda$4686 line.
In systems with a strong nebular continuum component, $T_e$ can also be constrained from the magnitude of the Balmer jump (e.g. \citealt{Guseva2006, PerezMontero2017}). We return to this in Section~\ref{sec:continuum} where we show that $T_e$(H$^+$) implied by the Balmer jump is consistent with $T_e$(O$^{++}$) measured from the \OIII\ auroral line ratio.

\subsection{Electron density}

The density-sensitive doublets of C~{\sc iii}] and [O~{\sc ii}] are not resolved in our observations, while N~{\sc iv}] and [S~{\sc ii}] are not detected. 
We report a marginal detection of the [Ar~{\sc iv}] $\lambda\lambda$4711, 4740 doublet in the Prism spectrum.
These lines are partially blended with each other and with He~{\sc ii} $\lambda$4686, while [Ar~{\sc iv}] $\lambda$4711 is also completely blended with He~{\sc i} $\lambda$4713. 
Our three-component fit to this complex yields [Ar~{\sc iv}]~$\lambda$4711/$\lambda4740=1.6\pm0.8$ after subtracting the predicted He~{\sc i} $\lambda$4713 contribution (assuming $\lambda$4713/$\lambda$4471 = 0.15), consistent with the low-density limit ($n_e\lesssim10^3$~cm$^{-3}$; \citealt{Kewley2019_dens}).
A consistent density constraint arises if the UV continuum turnover in \objA\ is driven by two-photon nebular continuum, since the feature strongly suppressed by $l$-changing collisions at higher densities.
The presence of the two-photon continuum will be discussed in Section~\ref{sec:continuum}.

\begin{table}
\caption{Physical properties and chemical abundances derived for \objA.}
\centering
\begin{tabular}{lll}
\hline
Property & Derived value  \\
\hline
$T_e({\rm O}^{++})$ & $18,300\pm1,500$ K  \\
$n_e$ & $\lesssim 10^3$ cm$^{-3}$ \\
\hline
$12+\log ({\rm O}^{++}/{\rm H}^+)$ & $7.58\pm0.01$\\
$12+\log ({\rm O}^{+}/{\rm H}^+)$ & $5.79^{+0.04}_{-0.03}$  \\
$12+\log ({\rm O}/{\rm H})$ & $7.59\pm0.01$ \\
\hline
$\log({\rm N}^{++}/{\rm O}^{++})$ & $< -0.85$  \\
$\log({\rm C}^{++}/{\rm O}^{++})$ & $-0.76\pm0.03$  \\
$\log({\rm C}/{\rm O})$ & $-0.73\pm0.03$  \\
$\log({\rm Ne}/{\rm O})$ & $-0.85^{+0.03}_{-0.04}$  \\
\hline
${\rm He}^{+}/{\rm H}^+$ & $0.10\pm0.01$  \\
${\rm He}^{++}/{\rm H}^+$ & $0.005\pm0.001$  \\
${\rm He}/{\rm H}$ & $0.11\pm0.01$ \\
\hline
$\log({\rm O}^{++}/{\rm O}^+)$ & $1.79^{+0.03}_{-0.04}$  \\
${\rm He}^{++}/{\rm He}^+$ & $0.06\pm0.01$  \\
\hline
\end{tabular}
\label{tab:physical_properties}
\end{table}

\subsection{Chemical abundances}
\label{sub:chem_abun}
We derive chemical abundances for \objA\ adopting $T_e=1.83\times10^4$~K and a density of $n_e=100$ cm$^{-3}$ following the procedure in \citet{Cameron2023_GNz11}. Given the apparent agreement between $T_e({\rm H}^+)$ and $T_e({\rm O}^{++})$, we assume a constant temperature for all ionisation zones. We derive $12+\log({\rm O}/ {\rm H})=7.59\pm0.01$ from the \OII, \OIII, and H$\beta$ lines, obtaining a consistent result with both Prism and grating line fluxes.
We derive the carbon abundance from the C~{\sc iii}]~$\lambda\lambda$1907,~1909~/~[O~{\sc iii}]~$\lambda$5007 ratio measured from the Prism, assuming no dust, finding $\log ({\rm C}/{\rm O})=-0.73\pm0.03$ after applying the ionisation correction factor (ICF) presented in \citet{Amayo2021}. However, we note the significant emission from C~{\sc iv} in the spectrum. 
The ICF may not be representative of the extreme conditions in \objA\ (e.g. \citealt{Berg2019}), and the quoted C/O may represent a lower limit.

Since no nitrogen lines are detected in \objA, we can only place upper limits on the nitrogen abundances.
The low O$^{+}$ abundance implies that the N$^{+}$ abundance is likely negligible.
We instead consider the N$^{++}$ abundance using N~{\sc iii}] $\lambda\lambda$1750 / [O~{\sc iii}] $\lambda$5007 limit measured from the Prism, and the N~{\sc iii}] $\lambda\lambda$1750 / O~{\sc iii}] $\lambda$1666 limit measured from the grating. These yield 3-$\sigma$ upper limits of $\log({\rm N}^{++}/{\rm O}^{++})<-0.85$ and $<-1.01$ respectively. We adopt the former as our preferred limit.

We detect several helium recombination lines from both the singly and doubly ionised states. Prism measurements of He~{\sc i} $\lambda$4471 / H$\beta$ and He~{\sc i} $\lambda$5875 / H$\beta$ yield consistent measurements of ${\rm He}^{+}/{\rm H}^{+}=0.10\pm0.005$ and $0.10\pm0.01$ respectively. Deriving a ${\rm He}^{++}/{\rm H}^{+}$ abundance using the He~{\sc ii} $\lambda$4686 line results in only a $6\pm1$~\% contribution to the total helium abundance, giving ${\rm He}^{++}/{\rm H}^{+}=0.005\pm0.001$.
Together, this implies a total helium abundance of ${\rm He}/{\rm H}=0.11\pm0.01$, higher than typical values observed in low-metallicity systems \citep{Matsumoto2022_HeAbun}, but consistent with some massive globular clusters \citep{Piotto2007}. 
However, we caution that our derived helium abundance 
does not account for collisional emission or self-absorption, which could be significant (e.g. \citealt{Peimbert1992}). 

The values obtained from these measurements are summarised in Table~\ref{tab:physical_properties}.

\subsection{Photoionization Modelling of the Emission Lines}
\begin{figure*}
    \centering
    \includegraphics[width=0.95\textwidth]{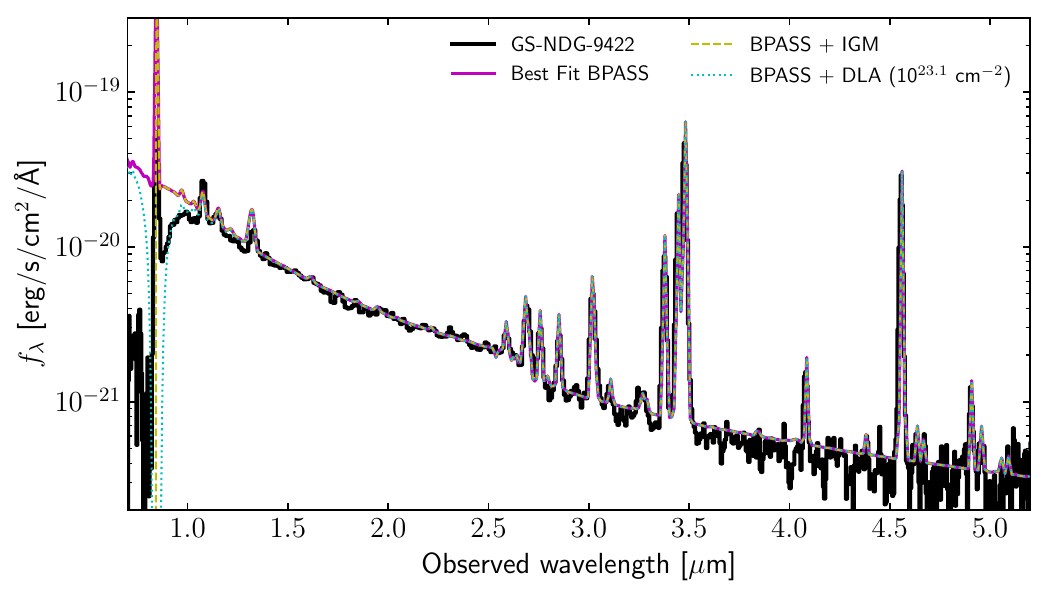}
    \caption{Prism spectrum of \objA\ (black) compared with the best fit models using a standard SSP (magenta) accounting for both the $z=6$ IGM opacity (dashed yellow) and a DLA with column density of $10^{23.1}\ {\rm cm^{-2}}$ (dotted cyan).}
    \label{fig:best_fit_bpass_full}
\end{figure*}

The diagnostic diagrams, electron temperature, and line widths all point to a scenario where \objA\ is powered by emission from young stars. We explore the feasibility of reproducing the emission lines with young stellar populations with photoionization models using {\sc CLOUDY} v23 \citep{Ferland2017}. We assume that the intrinsic spectrum is powered by a standard young SSP model, adopting {\sc BPASS v2.2.1} SSP models including binary stellar evolution \citep{Eldridge2017} with an IMF maximum mass of 300~M$_{\odot}$ and a high-mass slope of $-2.35$. 
We consider a population with an age of 3~Myr and a metallicity of 0.1$Z_{\odot}$, consistent with that measured for the gas from the spectrum. 
We note that the abundance patterns assumed in these models are scaled solar, which may not be representative of the stellar populations forming at this redshift.
We adopt a spherical geometry with an inner radius of $0.1$~pc. The calculation is stopped at an electron fraction of 1\% or when the neutral column density reaches $10^{18.7}\ {\rm cm^{-2}}$, consistent with the minimal observed Ly$\alpha$ emission offset ($\lesssim100$~km~s$^{-1}$) \citep{Verhamme2015}. We then vary gas density, ionization parameter, metallicity (assuming solar abundance patterns from \citealt{Grevesse2010}), and carbon abundance, fixing the gas temperature to that measured from [O~{\sc iii}]~$\lambda$4363/$\lambda$5007, until we reproduce the line strengths of [O~{\sc iii}]~$\lambda$5007, [O~{\sc ii}]~$\lambda\lambda$3727, C~{\sc iii}]~$\lambda\lambda$1909, H$\alpha$, and H$\gamma$ with respect to H$\beta$. Fe is assumed to be heavily depleted or to have not yet been produced. We also assume resonant lines from low-ionization species have the same escape fraction as that measured for Ly$\alpha$. Note that by focusing on line ratios, the resulting continuum is a prediction of the model. The intrinsic spectrum of our best fit model ($n=10^{3}\ {\rm cm^{-3}}$, $\log_{10}(U)=1.2$, $\log_{10}(Z_{\rm O}/Z_{\odot})=-1.1$, and $\log_{10}(Z_{\rm C}/Z_{\odot})=-1.4$) is shown as the magenta line in Figure~\ref{fig:best_fit_bpass_full}.

As can be seen in Figure~\ref{fig:best_fit_bpass_full}, the strengths of the emission lines are well reproduced by our best fit {\small CLOUDY} model, highlighting the fact that relatively metal-poor, young stellar populations are able to explain the emission lines seen in this galaxy. There remains a discrepancy between the observed He~{\sc ii} emission and that predicted by the photoionization models\footnote{For example, the He~{\sc ii}~$\lambda$1640/H$\beta$ ratio is measured to be $0.26\pm0.05$ whereas the photoionization models predict a value of 0.03.}, but this is consistent with numerous low-redshift metal-poor galaxies that show anomalous He~{\sc ii} emission with no signatures of a black hole or an AGN \citep{Kehrig2015, Kehrig2018, Schaerer2019, Senchyna2020, Saxena2020_HeII}.


\begin{figure}
    \centering
    \includegraphics[width=0.45\textwidth]{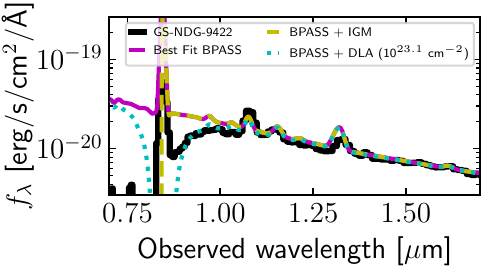}
    \includegraphics[width=0.45\textwidth]{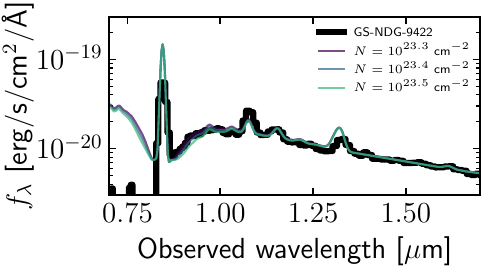}
    \includegraphics[width=0.45\textwidth]{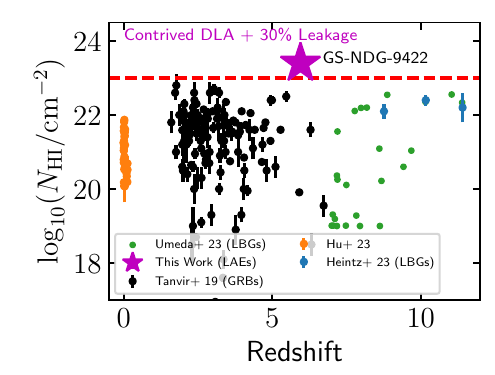}
    \caption{
    {\it Top:} Zoom-in on UV turnover for models plotted in Figure~\ref{fig:best_fit_bpass_full}.
    {\it Middle:} Fitting to \objA\ using a DLA model with a 70 \% covering fraction allows Ly$\alpha$ escape, but implies an extremely high column density.
    {\it Bottom:} Comparison of implied column density with known DLAs.
    }
    \label{fig:bpass_with_dla}
\end{figure}

\begin{figure}
    \centering
    \includegraphics[width=0.5\textwidth]{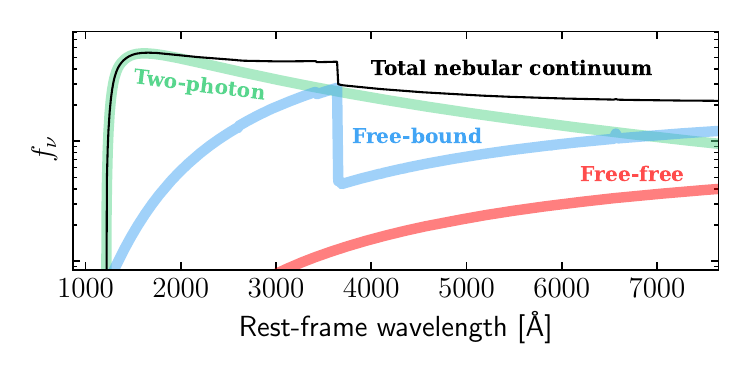}
    \caption{
    Schematic of the main nebular continuum components across the rest UV to optical range. Free-bound emission can give rise to a spectral discontinuity at the Balmer limit ($\lambda_{\rm rest}\approx3645$ \AA{}). Two-photon continuum has a fixed shape with a turnover at $\lambda_{\rm rest}\approx1430$ \AA{}, but is usually subdominant compared to continuum emission of the ionising source. Free-free emission is usually sub-dominant at these wavelengths. The relative contribution of two-photon continuum is highly dependent on nebular conditions.
    }
    \label{fig:nebular_continuum}
\end{figure}

\section{Continuum shape} \label{sec:continuum}
We have shown in the previous section that the emission lines are consistent with ionisation by young metal-poor stellar populations. In this section, we explore the possible origin of the continuum emission in \objA.

\subsection{Balmer jump}

Low-metallicity, young stellar populations can have ionizing photon production efficiencies that are high enough that the free-bound emission from recombining hydrogen can outshine the stellar continuum at optical wavelengths, leading to the observation of a Balmer jump \citep[e.g.][]{Byler2017}.
As seen in Figure~\ref{fig:spectrum}, already from the broad- and medium-band photometry alone, there is evidence of a spectral discontinuity at the location of the Balmer limit. The Balmer jump is not typically observed as a sharp spectral feature since, at realistic spectral resolutions, blending of the tail of the Balmer series and strong emission lines like \OII\ $\lambda\lambda$3726, 3729 and [Ne {\sc iii}] $\lambda$3869 can occur \citep[e.g.][]{Schirmer2016_NEBULAR}. We perform an empirical fit for this discontinuity by masking out regions contaminated by strong emission lines and fitting the spectrum (in units of $f_\nu$) over the range $\lambda_{\rm rest}>1930$ \AA{} with a two-part linear function, broken at the Balmer limit ($\lambda_{\rm rest}=3645$ \AA{}). This results in a clear detection of a spectral jump with $15.0\pm0.9$~nJy in the observed frame (second panel Figure~\ref{fig:spectrum}; dotted line), demonstrating the strong contribution of nebular continuum to the spectrum of \objA. 
Such features have been observed in the spectra of metal-poor star-forming galaxies at low-redshift \citep{Guseva2006, Guseva2007} and at high-redshift \citep{RobertsBorsani2024}, while they have also been widely predicted at high-redshift in SED modelling of photometric data \citep{Endsley2023, Topping2023} and in simulations \citep{Katz2023_SPHINX_DR, Wilkins2023}.

As shown in Figure~\ref{fig:best_fit_bpass_full} the spectral discontinuity at the location of the Balmer jump is well reproduced by our best fit photoionization model. It is important to emphasize here that the presence of spectral discontinuity in the continuum is a prediction of the photoionization model which was purely designed to reproduce the emission lines rather than the shape of the continuum. 

\subsection{UV continuum turnover}
Even more striking than the spectral discontinuity at the location of the Balmer jump is the presence of a steep turnover in the UV continuum of \objA\ at $\lambda_{\rm obs}\approx1$~$\mu$m ($\lambda_{\rm rest}\approx1430$~\AA{}) (Figure~\ref{fig:spectrum}). While our best-fit photoionization model is successful in reproducing the line emission and the Balmer jump, where the model fails is that it significantly over-predicts the emission at $\lambda_{\rm rest}\lesssim1430$~\r{A}. Here we explore the origin of this discrepancy.

\subsubsection{Absorption from neutral hydrogen?}
\label{sub:dla}

Similar UV turnovers are often observed as a result of absorption from foreground neutral hydrogen – either the neutral intergalactic medium (IGM) \citep{Jordi1998} or Damped Lyman-$\alpha$ absorption (DLA) systems \citep{Heintz2023_DLA}.

Beginning with the IGM, we apply the $z=6$ IGM transmission curves from \citet{Garel2021}, to our best-fit photoionization model and we find that IGM damping is unable to produce the observed UV turnover from this BPASS model (yellow dashed line in Figures~\ref{fig:best_fit_bpass_full}~\&~\ref{fig:bpass_with_dla}). Indeed, the observed turnover in \objA\ requires absorption far exceeding the maximal IGM damping wing, calculated following the formalism presented in \citet{Jordi1998} and \citet{Barkana2001}. Similar results were found in \citet{Heintz2023_DLA} for targets at much higher redshift. This is not particularly surprising because if the IGM were responsible, many more galaxies at $z\gtrsim6$ would show very strong UV turnovers.

We next consider the presence of a DLA. We model damping due to the presence of a DLA with {\sc CLOUDY} by calculating the transmission of the best-fit BPASS model through slabs of neutral hydrogen with increasing column densities, finding that column densities of $N_{\rm H}\sim10^{23}$~cm$^{-2}$ are needed to reproduce the magnitude of the turnover. Such a value is higher than any previously reported galaxy-scale DLA system (e.g. \citealt{Tanvir2019, Umeda2023}; see Figure~\ref{fig:bpass_with_dla}).

The primary issue with naively assuming a DLA is that such high column densities imply zero transmission at 1216~\AA{}, conflicting with the strong observed Ly$\alpha$ emission. The middle panel of Figure~\ref{fig:bpass_with_dla} shows that this can, in principle, be reconciled by invoking a DLA with 30~\% leakage. However, the plausibility of such an extreme column density with a low covering fraction is unclear, given the fact that other known DLA systems with very high gas columns do not show Ly$\alpha$ emission \citep{Umeda2023,Heintz2023_DLA}. 
In principle, one could shift the DLA to a lower redshift which would allow for some Ly$\alpha$ emission to escape as the emitted Ly$\alpha$ will be redshifted out of resonance in the reference frame of the DLA; however, this would require much higher column densities than $10^{23}$~cm$^{-2}$ in order to reproduce the UV turnover. Furthermore, at such high column densities, the gas can self-shield from the local-radiation field, and the core would be expected to be fully molecular. It is not clear whether such high gas columns can be maintained without the gas collapsing and forming stars \citep{Schaye2001}. Moreover, \objA\ shows no signatures of dust extinction based on the Balmer decrement, the He~{\sc II} ratio, and the fact that the photoionization model can easily reproduce the observed UV slope without assuming dust (Section~\ref{sec:physical_conditions}). If the DLA was present within \objA\ at a metallicity of nearly 10\% solar, one would need to explain why the dust has not formed or could not survive in a thick neutral cloud. Given the fine-tuned requirements, we consider this picket-fence scenario of a thick DLA with optically-thin channels highly unlikely. 

Other geometries may exist (apart from the picket fence or lower-redshift DLA) that could possibly explain \objA. For example, one could consider a foreground DLA and background or spatially offset clouds where emission could be reflected. These scenarios all must be reconciled with the very high Ly$\alpha$ escape fraction of $\sim27\%$ which again seems unlikely. For this reason, we consider other alternatives for the origin of the UV turnover.

~\\

\subsubsection{Two-photon continuum emission}
\label{sub:neb_cont_fit}
The nebular continuum consists of three components: free-bound, which gives rise to the spectral discontinuity at $\lambda_{\rm rest}\sim3645$~\AA{}, free-free, which is typically subdominant compared to free-bound and only impacts $\lambda_{\rm rest}\gtrsim3000$~\AA{}, and two-photon emission (Figure~\ref{fig:nebular_continuum}). 
Two-photon emission arises from transitions from $2s\rightarrow1s$ in neutral hydrogen, resulting in the emission of two photons whose energies sum to that of Ly$\alpha$.
The distribution of photons emitted via this process is symmetric around 2431 \AA{} ($\nu = \frac{1}{2}\nu_{\rm Ly\alpha}$) when expressed in terms of number of photons per second per frequency interval. However, when expressed in units $f_\lambda$,
it takes the form of a broad asymmetric peak which turns over at $\lambda_{\rm rest}\approx1430$~\AA{}, remarkably close to the wavelength of the observed UV turnover in \objA. The two-photon continuum is typically subdominant compared to the stellar continuum, but has been predicted to be observable in systems with extremely high ionising photon production efficiency \citep{Fosbury2003, Raiter2010}.
Here we consider the possibility that the observed UV turnover is the two-photon continuum.

To test whether the continuum of \objA\ is consistent with being primarily nebular, we consider a model where the spectrum consists of:
\begin{enumerate}
\item Two-photon emission
\item Free-bound \& Free-free emission
\item Young stars with ages $<10^{7.5}$~yrs
\item Old stars with ages $>10^{7.5}$~yrs
\end{enumerate}

~\\

The shape of the spontaneous two-photon continuum is fixed\footnote{A significant contribution of induced two-photon emission can alter the width of the probability distribution of emitted photons \citep[e.g.][]{Chluba2006}, which can in turn shift the wavelength at which the two-photon continuum turns over when expressed in $f_\lambda$. However, we found this effect to be completely negligible ($<1$ \AA{}) in any of the modelling considered in this work.}, so we only vary its overall normalization. The combination of free-bound and free-free emission has a shape that is sensitive to gas temperature and thus we vary both gas temperature and the normalization of this component. Finally the shapes of the stellar spectra are sensitive to age, so we consider the normalizations and ages of each stellar population. In total, the model has seven free parameters. As above, we assume that the stellar component follows {\sc BPASS v2.2.1}, while the nebular continuum components are computed with {\small PYNEB} \citep{Luridiana2015_pyneb}. Posterior distributions for each parameter are computed using an MCMC \citep{emcee2013PASP}.

\begin{figure}
    \centering
    \includegraphics[width=0.49\textwidth]{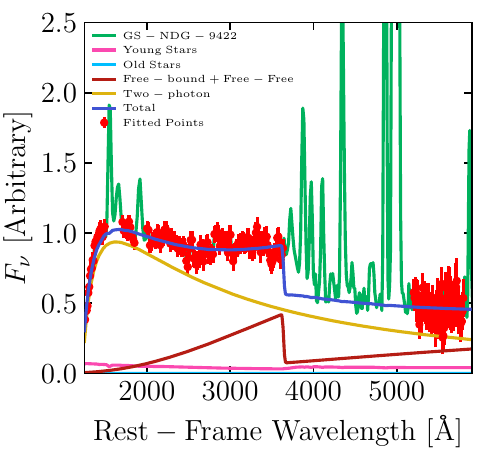}
    \includegraphics[width=0.49\textwidth]{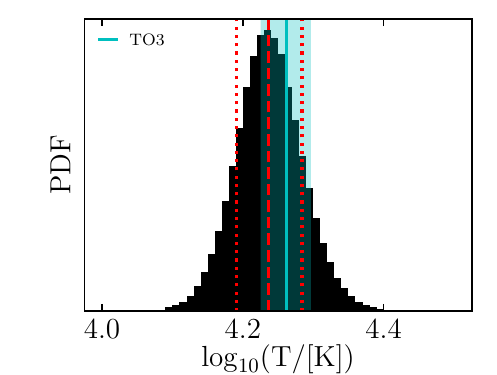}
    \caption{
    {\it Top:} Best fit continuum model from fitting described in Section~\ref{sub:neb_cont_fit}. Continuum points using in the fitting are shown in red, while the full spectrum is shown in green. Coloured lines show the continuum components arising from young stars (pink), old stars (light blue), free-bound + free-free nebular (brown), and two-photon (yellow) in the best fit model. The blue line shows the overall best fit.
    {\it Bottom:} Posterior distribution on gas temperature predicted by the continuum fitting. The median and 1$\sigma$ values (red lines) are in good agreement with the nebular temperature measured from the \OIIIl4363/$\lambda$5007 ratio (Table~\ref{tab:physical_properties}; blue).
    }
    \label{fig:neb_cont_fitting}
\end{figure}

\begin{figure}
    \centering
    \includegraphics[width=0.49\textwidth]{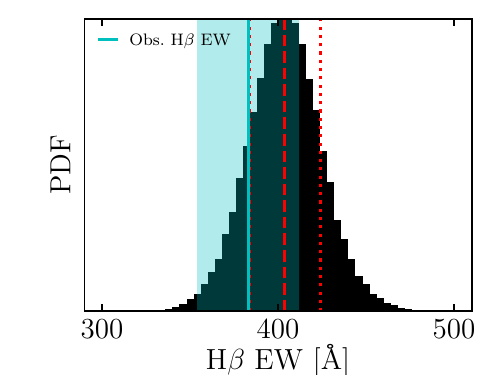}
    \includegraphics[width=0.49\textwidth]{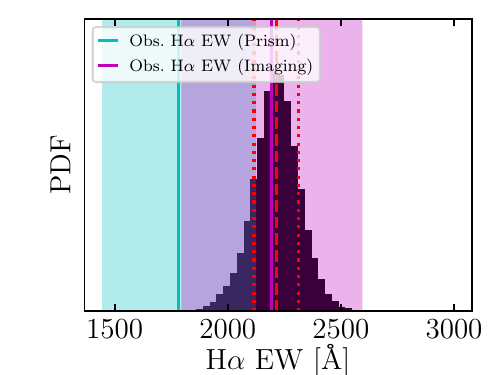}
    \caption{
    {\it Top:} Posterior distribution on H$\beta$ equivalent width predicted by the continuum fitting. The median and 1$\sigma$ values (red lines) are consistent with that measured from the Prism spectrum (blue).
    {\it Bottom:} Same as top panel, but for H$\alpha$. Here, the median and 1$\sigma$ values (red lines) are somewhat higher than that measured from the Prism spectrum (blue), but in good agreement with the equivalent width implied by the medium-band imaging.
    }
    \label{fig:neb_cont_fitting_ha}
\end{figure}

In the top panel of Figure~\ref{fig:neb_cont_fitting} we show the model corresponding to the 50th percentile distribution of each of the seven parameters (blue line) as well as the points used for the fit (red) and the observed spectrum (green). The MCMC prefers a model where the continuum at $\lambda_{\rm rest}<5800$~\r{A} is dominated by the nebular continuum. Indeed the two-photon continuum is able to reproduce of the shape of the UV downturn while free-bound emission peaks high enough to create the observed spectral discontinuity near the Balmer jump. Applying frequentest statistics, we find a reduced $\chi^2$ of $\chi^2_{\nu}=0.80$ indicating that this 50th percentile model is a very good fit to the continuum of \objA.

Although this nebular-dominated scenario provides a good fit to the continuum, it remains an open question of whether the model is consistent with the observed emission lines. This can be verified in two ways. Because the free-free and free-bound emission are sensitive to gas temperature, we can check whether the temperature predicted from the continuum is consistent with that measured from the oxygen emission lines. In the bottom panel of Figure~\ref{fig:neb_cont_fitting}, we show the posterior distribution on gas temperature predicted by the continuum fitting (black histogram) compared to that measured using the [O~{\small III}]~$\lambda$4363 auroral line (cyan), finding that the two are consistent within $1\sigma$ uncertainty.
While formally the O~{\sc iii} temperature does not have to be the same as that of the H~{\small II} gas, empirical measurements from low-redshift galaxies suggest that the two temperatures are often very similar \citep[e.g.][]{Guseva2006,Guseva2007}.

An even stronger test is to compare the observed H$\beta$ equivalent width versus that predicted by the continuum-fitting model. 
The H$\beta$ emission should primarily arise from recombination, the same as the free-bound continuum. Thus, in a nebular-dominated scenario, EW(H$\beta$) should only depend on gas temperature and the relative contribution of free-bound and free-free to two-photon emission. 
The top panel of Figure~\ref{fig:neb_cont_fitting_ha} shows that the $1\sigma$ uncertainty on the observed EW(H$\beta$) significantly overlaps the $1\sigma$ spread in the posterior distribution of predicted EW(H$\beta$). This again demonstrates that the information in the continuum is sufficient to explain many of the properties of the emission lines.

We can apply the same test for H$\alpha$ emission. Since the continuum was only fit up to rest-frame 5800~\r{A}, predicting the H$\alpha$ equivalent width requires extrapolating the model.
In Figure~\ref{fig:neb_cont_fitting_ha} we compare the posterior distribution on predicted EW(H$\alpha$) to that measured from the prism as well as that inferred from imaging. As discussed above, the EW$_0$(H$\beta$) from the imaging data is somewhat higher than that measured from the prism (although the $1\sigma$ contours overlap). We find that our predicted values fall high compared to the prism measurement, but the value from the imaging falls on top of our $1\sigma$ confidence interval. Hence our model is formally very consistent with the imaging data and consistent within $2\sigma$ of the prism. Because we have not fit the continuum near H$\alpha$ the model could be missing a contribution from older stars, which can increase at these wavelengths without impacting our current fit. Since \objA{} has metals, these must have originated somewhere. A metallicity of $0.1~Z_{\odot}$ is much higher than that predicted for the IGM at $z\sim6$ \citep[e.g.][]{Madau2014} and thus it is unlikely the system was enriched externally. While it is possible that we are witnessing the illumination of the immediate enrichment from the current population of massive stars, the abundance patterns are such that it is likely that there might be an underlying population of stars that is no longer UV-bright. Therefore our current model remains flexible enough to accommodate the scenario of an older population of stars that contributes to the continuum at wavelengths near H$\alpha$.

Nevertheless, given that our model prefers that the UV and optical part of the spectrum arises primarily from nebular emission, it poses the question: what drives this behaviour?


\begin{figure}
    \centering
    \includegraphics[width=0.45\textwidth]{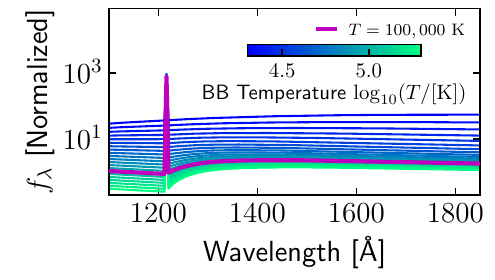}
    \includegraphics[width=0.45\textwidth]{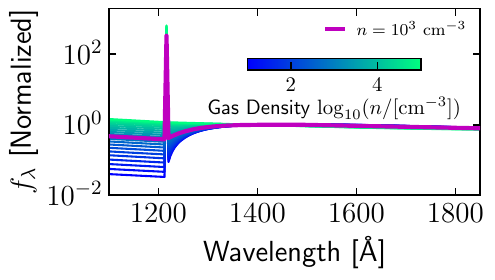}
    \caption{{\it Top:} Normalized spectrum of hydrogen-only gas with $n=10^3$ cm$^{-3}$ irradiated by blackbodies of different temperatures (as given in the colour bar). The magenta line represents a blackbody temperature of 100,000~K. 
    {\it Bottom:} Normalized spectrum of hydrogen-only gas irradiated by a blackbody with $T=10^5$~K at different gas densities (as given in the colour bar). The magenta line represents a density of $10^3\ {\rm cm^{-3}}$.}
    \label{fig:bb_t_vary}
\end{figure}

\begin{figure*}
    \centering
    \includegraphics[width=0.95\textwidth]{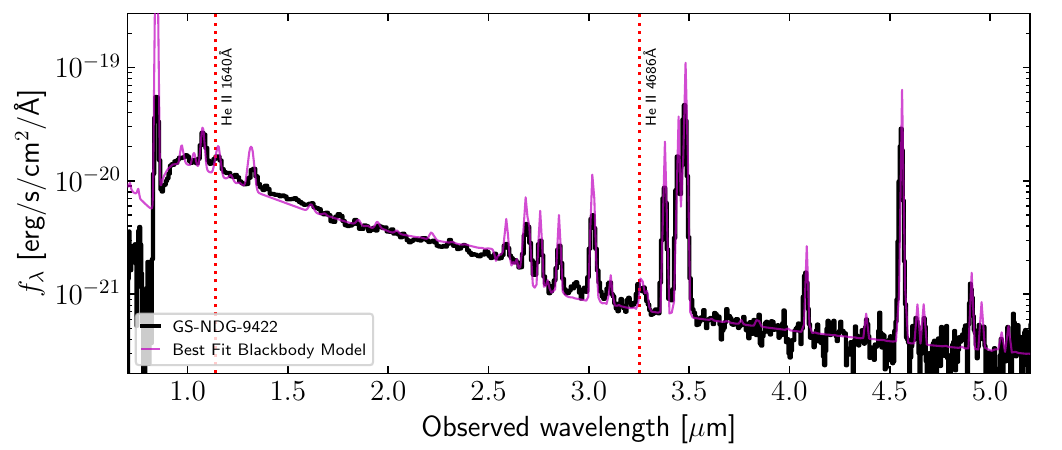}
    \includegraphics[width=0.95\textwidth]{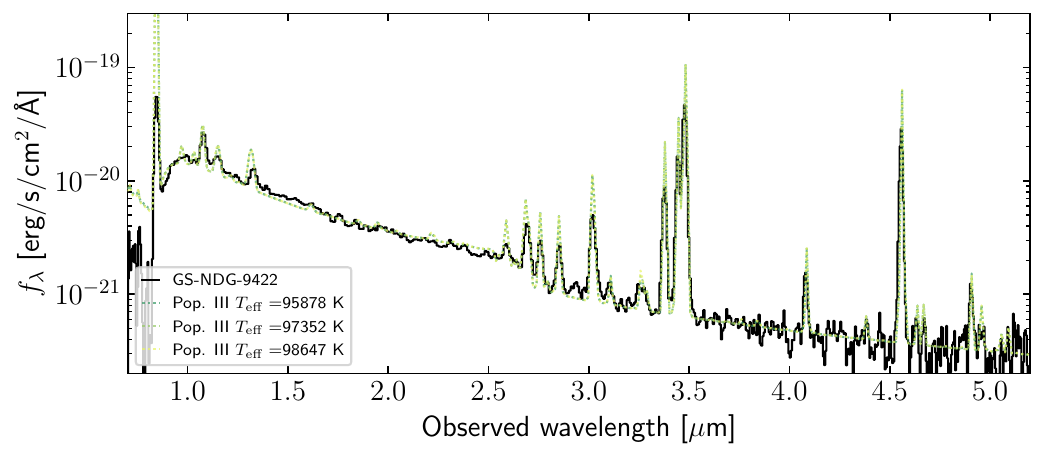}
    \includegraphics[width=0.95\textwidth]{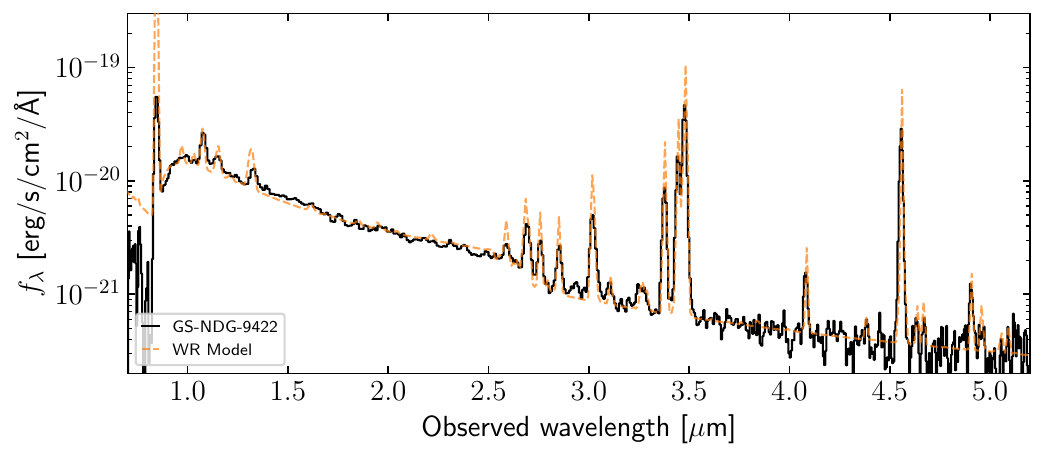}
    \caption{
    {\it Top:} Spectrum of \objA\ (black) compared with the best fit blackbody model (magenta) with $T_{\rm BB}=10^{5.05}$ K and a He~{\sc ii} leakage fraction of 0.22 (Table~\ref{tab:bb_params}; Section~\ref{sub:blackbody_model}).
    {\it Middle:} Spectrum of \objA\ (black) compared with photoionsiation models powered by various individual massive Pop.~III stars with different effective temperatures. {\it Bottom} Spectrum of \objA\ (black) compared with the photoionisation model powered by a model Wolf-Rayet SED with the most similar spectrum to the best-fit blackbody SED. All models can also reproduce the observed spectrum well (dotted lines). See Section~\ref{sub:blackbody_model} for details.
    }
    \label{fig:best_fit_blackbody}
\end{figure*}

\begin{figure}
    \centering
    \includegraphics[trim={0cm 0cm 0cm 5.8cm},clip,width=0.45\textwidth]{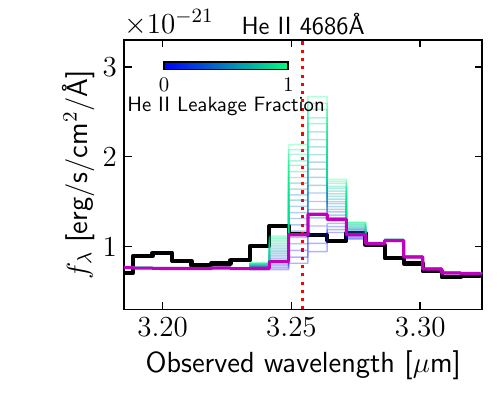}
    \caption{
    Zoom in on the region around He~{\sc ii}~$\lambda$4686 demonstrating the need to reduce the He~{\sc ii} ionizing flux in the blackbody models. The spectrum of \objA\ is shown in black. The magenta line shows the best-fit model from Figure~\ref{fig:best_fit_blackbody}. Green to blue lines show decrease He~{\sc ii} leakage fraction applied to the blackbody, modelling the effect of a He-rich atmosphere. Models with a high leakage fraction clearly overpredict the He~{\sc ii}~$\lambda$4686 line, although we note this feature is blended with nearby [Ar~{\sc iv}] lines.
    We note a slight flux excess blue-ward of He~{\sc ii}~$\lambda$4686 which, from inspection of the 2D spectrum, appears to be driven by a noisy pixel. This excess flux is still below the level of the excess emission implied by the 100~\% He~{\sc ii} leakage model shown in green.
    }
    \label{fig:best_fit_blackbody_HeII}
\end{figure}

\section{What could be driving two-photon emission in \objA?} 
\label{sec:spectral_modelling}
Under the assumption that the continuum of \objA~is nebular-dominated, we consider the numerous scenarios that may give rise to the observed spectrum.

\subsection{Is the ionising source present in the spectrum?}
\label{sub:geometry}

Although Balmer jumps have commonly been observed in the spectra of young, low-metallicity star-forming galaxies \citep[e.g.][]{Guseva2006,Guseva2007}, the steep UV slopes associated with the spectra of these stellar populations means that the nebular contribution is completely sub-dominant in the FUV where the two-photon continuum peaks. However, one can envision a scenario where the nebular emission is offset from the location of ionising source. If the slit were to contain only nebular gas, the two-photon emission could dominate the observed spectrum.

One example of this is Hanny's Voorwerp, which has been suggested to be quasar light echo \citep{Lintott2009}. In this case, both the free-bound and two-photon emission are expected to be strong. However, there are three key differences between Hanny's Voorwerp and \objA: (1) He~{\sc ii}~$\lambda$4686/H$\beta$ is $>6\times$ higher in the presumed QSO light echo compared to the galaxy studied here (Figure~\ref{fig:diagnostic_diagrams}). He~{\sc ii}~$\lambda$1640 is also much stronger in Hanny's Voorwerp \citep{Keel2012}, (2) while the two-photon continuum is likely important in Hanny's Voorwerp, another unidentified source has been postulated to explain the excess UV emission. In other words, the spectrum is not fully nebular dominated in the UV
and may consist of additional scattered light, and (3) the spatial extent of Hanny's Voorwerp is tens of kpc. 
Conversely, \objA\ is very compact, and it is well-centred within the NIRSpec micro-shutter (Figure~\ref{fig:spectrum}).
While other QSO light echoes have been detected \citep[e.g.][]{Schirmer2016b}, the spectra and morphologies do not seem to be consistent with \objA, leading us to disfavour this scenario.

An alternative explanation is that the ionization source has flickered on and off and we are capturing the system just after it has shut off. In this case, no continuum from the ionising source would be detected and we would be observing only the remnant nebular emission. However, the H$^+$ recombination timescale, assuming Case~B recombination at a temperature of 18,300~K, is $\sim5,000$~yr for a density of $10^2\ {\rm cm^{-3}}$ or $\sim500$~yr for $10^3\ {\rm cm^{-3}}$. This is much shorter than the main-sequence lifetimes of massive stars. Thus, in the case of flickering, it is highly unlikely that stars are powering the emission. Some QSO proximity zones show evidence for QSO lifetimes of $<10^4$~yr \citep{Eilers2018,Eilers2021}. While such QSOs are a rare subset of the general population, these lifetimes are broadly consistent with the recombination timescale. Because these QSOs are detected via their near zone, the QSOs are currently ``on'' and thus the UV continuum is still dominated by the AGN and not the nebular emission. Evidence for AGN fading on such short time-scales has also been observed locally \citep[e.g.][]{French2023}, but the spectra of these objects, in particular the low ionization state lines is inconsistent with \objA. Moreover, we have only considered the hydrogen recombination time scale. If we consider He$^{++}$ under the same assumptions, we arrive at a recombination timescale of $\sim300$~yr for a density of $10^2\ {\rm cm^{-3}}$ or $\sim30$~yr for $10^3\ {\rm cm^{-3}}$. 
The fact that we observe He~{\sc ii} emission is thus difficult to reconcile with this `flickering' scenario, and imply that we would be catching the source at a very specific moment time.
Given the very special timing required, the identification of other objects with spectra like \objA\ would seemingly disfavour this scenario. We discuss the identification of similar candidates in Section~\ref{sub:other_objects}.

\subsection{Is \objA{} powered by hot stars?}
\label{sub:blackbody_model}
Assuming that the ionizing source remains luminous and is present within the slit, in order for both the two-photon and free-bound continua to dominate over the stellar spectrum in the rest-frame UV and optical, the source population must have a much larger ionising photon production efficiency ($\xi_{\rm ion}$) than standard SSP models, necessitating hotter blackbody temperatures. We explore this by running {\sc CLOUDY} simulations with input blackbody SEDs, with a setup closely based on that described in Section~\ref{sub:dla}. 
To gain insight into the requirements for the two-photon continuum to dominate over the ionizing SED, we initialize hydrogen-only gas at constant gas temperature (measured from [O~{\sc iii}]~$\lambda$4363/$\lambda$5007) and systematically vary the density and blackbody temperature (Figure~\ref{fig:bb_t_vary}). A weak UV turnover begins to appear at blackbody temperatures $T_{\rm BB}\gtrsim65,000$~K, which is hotter than a typical O star. A strong UV turnover requires at least $T_{\rm BB}\gtrsim 90,000$~K, much hotter than massive O~stars (up to $\sim$50,000~K; \citealt{Walborn2004,Evans2011,Bressan2012}).

The two-photon continuum is also sensitive to gas density because, at high densities, $l$-changing collisions will suppress the two-photon emission relative to Ly$\alpha$.
This is seen in the bottom panel of Figure~\ref{fig:bb_t_vary} where we vary gas density for a fixed blackbody temperature of $100,000$~K. The UV turnover is strongly suppressed at $n\gtrsim 10^3\ {\rm cm^{-3}}$.
We emphasize that the details of this calculation are sensitive to the chosen column density at which the model is truncated (which in our case is set by the velocity offset of Ly$\alpha$). Furthermore, there exist significant differences in atomic data predictions for the strength of $l$-changing collisions as a function of temperature \citep{Guzman2017}. 

Under the constraints determined above, we adopt an empirical approach to determine the possible underlying spectrum of \objA. 
We assume that the ionizing spectrum, to first order, can be modelled as a blackbody. To optimize the blackbody model fit to \objA, we begin with the parameters of the best fit {\small BPASS} model (Section~\ref{sub:dla}) and update ionization parameter, gas density, and blackbody temperature to simultaneously reproduce the emission line ratios and continuum shape. We allow for both density and ionization bounded nebulae by adding an additional stopping criterion to reproduce the measured [O~{\sc iii}]~$\lambda$5007/[O~{\sc ii}]~$\lambda\lambda$3727 (O32) ratio.
This stopping criterion often supersedes the neutral gas column stopping criteria used above. 
The O32 ratio and the gas temperature are allowed to vary within their observational uncertainties. 
Optimisation of this model (top panel of Figure~\ref{fig:best_fit_blackbody}) demonstrates that a blackbody model with $T=10^{5.05}$~K can provide a good fit to both the shape of the continuum and the majority of lower-ionization state emission line ratios in \objA.

However, where our simple blackbody model fails is that it strongly \emph{overpredicts} the flux of He~{\sc ii} lines compared to those observed in the spectrum (Figure~\ref{fig:best_fit_blackbody_HeII}).
The weak He~{\sc ii}~$\lambda$1640 and He~{\sc ii}~$\lambda$4686 in \objA\ places tight constraints on the ionizing spectrum at $E>4$~Rydberg.
The opacity of helium in the stellar atmosphere is well known to play an important role in mediating the flux of stellar populations at these energies, suppressing the flux of He$^+$-ionising photons ($\lambda<228$ \AA{}; e.g. \citealt{Smith2002}).
To explore this, we run models varying the leakage fraction of photons with $E>4$~Rydberg from 0\% (i.e. no He$^+$-ionizing photons) to 100\% (unattenuated blackbody). We conclude that $70\%-75\%$ of the He$^+$-ionizing photons emitted by the blackbody must be extinguished to match the observed He~{\sc ii} emission, with a leakage fraction of $0.25$ in our best-fit model.
The parameters for our optimized model are reported in Table~\ref{tab:bb_params}.

We note that other sources, including active galactic nuclei (AGN) or high-mass X-ray binaries (HMXBs), can have significant high-energy photon outputs, and have been invoked to explain peculiar emission line ratios seen at high-redshift \citep{Maiolino2023,Katz2023}. 
However, the weak He~{\sc ii} emission disfavours power-law SEDs that extend past the He$^+$-ionizing edge. We exclude HMXBs due to the strong over-prediction of He~{\sc ii} emission in these models (see Appendix~\ref{app:xrbs}), while the presence of an AGN is discussed above.
Finally, we note that spectral fitting of some low-redshift `high-redshift analogues' has suggested a similar need for significant ionizing contribution from a hot ($T>80,000$~K) blackbody \citep{Olivier2022}, however the key difference in \objA\ presented here is the dominance of the two-photon continuum. This necessitates an extremely high $\xi_{\rm ion}$ that cannot be produced if a substantial fraction of the ionizing photons are emitted by typical OB stars.

We now turn to the question of what sort of stars have sufficiently high surface temperatures to reproduce our best-fit blackbody model with $T_{\rm BB} = 10^{5.05}$ K.

\begin{figure*}
    \centering
    \includegraphics[width=\textwidth]{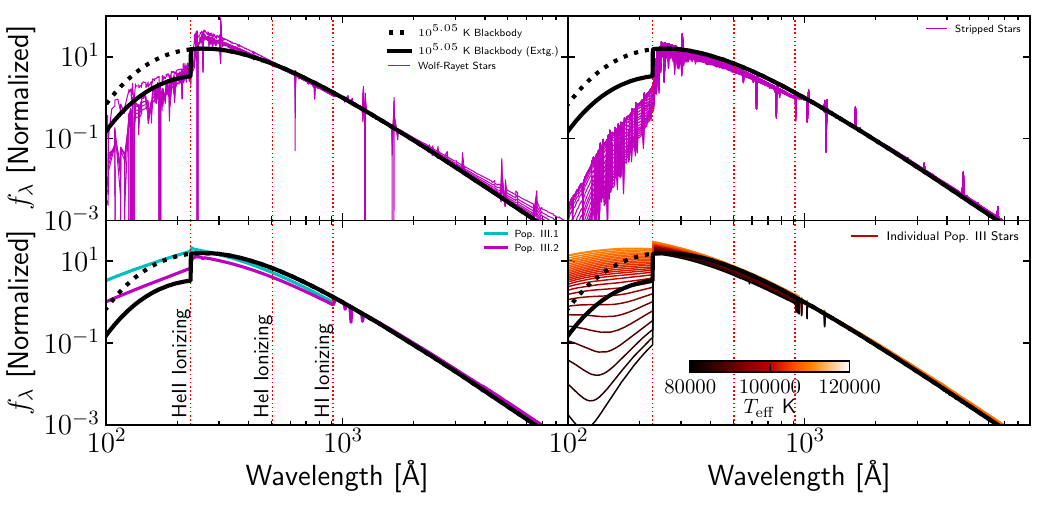}
    \caption{
    Best fit blackbody SED model with He-rich atmosphere (black solid) compared to hot star model SEDs from the literature. The dotted black line shows the blackbody spectrum prior to extinguishing the He~{\sc ii} ionizing radiation. The top left panel shows select $Z=0.07~Z_\odot$ Wolf-Rayet stars from \citet{Todt2015}, selected to match the measured nebular metallicity. These show remarkable resemblance to our model blackbody.
    The top right panel shows stripped star models from \citet{Gotberg2018}. These attain sufficiently high surface temperatures, but the atmospheres in existing models exhibit too much suppression of the He~{\sc ii} ionising continuum.
    The bottom left panel shows Yggdrasil Pop.~III models \citep{Zackrisson2011} with two different assumed Pop.~III IMFs (very top heavy, Pop.~III.1; blue) and (moderately top heavy, Pop.~III.2; magenta). These models strongly overpredict the He~{\sc ii}-ionising flux.
    The bottom right panel shows individual massive Pop.~III stars from \citet{Larkin2023} with different effective temperatures as shown in the colour bar. Some of these models exhibit the required SED properties to reproduce the observed spectrum of \objA\ (Figure~\ref{fig:best_fit_blackbody}), although we note the discrepancy between these zero-metallicity models and the measured nebular metallicity. Hence, we are not claiming the presence of Pop.~III stars in \objA.
    }
    \label{fig:spectral_overlay}
\end{figure*}

Wolf-Rayet stars not only exhibit extremely high surface temperatures, but can also have helium atmospheres that provide the necessary opacity to reduce their He$^{+}$ ionizing output \citep{Crowther2007}.
We explore models from grid of the PoWR Wolf-Rayet models \citep{Todt2015}\footnote{\url{https://www.astro.physik.uni-potsdam.de/~wrh/PoWR/powrgrid1.php}}. Given the gas-phase metallicity measured for \objA, we consider the WNL-H40 grid of the PoWR Wolf-Rayet models \citep{Todt2015} with $Z=0.07~Z_{\odot}$, similar to that measured for \objA. We identify model 13-10 as most similar to our optimised blackbody, for which $T=100,000$~K and the luminosity is fixed at $L=10^{5.3}L_{\odot}$ (Figure~\ref{fig:spectral_overlay} top left).
We note that this model assumes iron group abundances to be scaled solar, which may not be representative of stars forming at $z\sim6$. Abundances of carbon, nitrogen, and oxygen are assumed to have undergone significant CNO burning (see \citealt{Todt2015} for details).

Adopting parameters from the optimised blackbody model but removing the He~{\sc ii} leakage parameter, we replaced the blackbody SED with this theoretical low-metallicity Wolf-Rayet star spectrum. 
No further optimisation is performed and the resulting spectrum is shown as the dashed orange line in Figure~\ref{fig:best_fit_blackbody}, nearly identical to the optimised blackbody model.

Although Wolf-Rayet stars are typically associated with broad emission lines due to their high-velocity stellar winds, the absence of these features in \objA\ could simply be the result of the extremely bright nebular component outshining these wind features, as predicted by our photoionization models. Furthermore, in the absence of iron, which is the dominant source of stellar atmospheric opacity, wind speeds can drop below 500~km~s$^{-1}$, even at solar oxygen abundance \citep{Grafener2008}. Hence, high-redshift Wolf-Rayet dominated galaxies may not show broad He~{\sc ii}~$\lambda$1640 and He~{\sc ii}~$\lambda$4686 lines \citep{Grafener2015}.

Stars stripped in binaries can also exhibit the required surface temperatures above $10^5$~K \citep{Gotberg2018, Gotberg2023}. Compared to our best-fit extinguished blackbody, we find that stripped star SEDs \citep{Gotberg2018} produce too few He$^{+}$ ionizing photons, and the He~{\sc ii} emission is underpredicted by these stars (Figure~\ref{fig:spectral_overlay} top right).

Massive metal-free Population~III stars are predicted to have sufficiently high temperatures to power a two-photon dominated spectrum \citep{Schaerer2002,Trussler2023}. 
While the IMF of Population~III stars is highly uncertain \citep{Klessen2023}, comparing extremely top-heavy (Pop.~III.1) and moderately top-heavy (Pop.~III.2) Yggdrasil models \citep{Zackrisson2011} with our extinguished blackbody (Figure~\ref{fig:spectral_overlay} bottom left) indicates these produce too many He$^{+}$ ionizing photons, while predictions for even less top-heavy IMFs begin to fall short of the required $\xi_{\rm ion}$. In contrast, some individual Pop.~III star models from \citet{Larkin2023} with effective temperatures of $\sim97,000$~K and masses of 85~M$_{\odot}$ to 108~M$_{\odot}$ reasonably reproduce the required ionizing SED (Figure~\ref{fig:spectral_overlay} bottom right). 
We strongly emphasize that the measured metallicity of $12+\log_{10}({\rm O/H})=7.59\pm0.01$ suggests \emph{the presence of Pop.~III stars is highly unlikely}, unless we are witnessing the immediate enrichment and illumination of metals produced by primordial stars. Hence, we are not advocating that \objA{} hosts a population of primordial stars.
Nevertheless, stellar atmosphere models have large uncertainties at low metallicity, and few such models exist in the literature, so we consider these Pop.~III star models in our analysis under the assumption that the atmospheres may be representative of hot massive stars at very low metallicity.
Repeating the exercise described for our Wolf-Rayet models, this time replacing the blackbody SED with low-metallicity massive star models from \citet{Larkin2023} with effective temperatures between 95,000~K and 99,000~K also results in a very good fit to the spectrum of \objA\ (green dashed lines in Figure~\ref{fig:best_fit_blackbody}).

\begin{figure*}
    \centering
    \includegraphics[width=\textwidth]{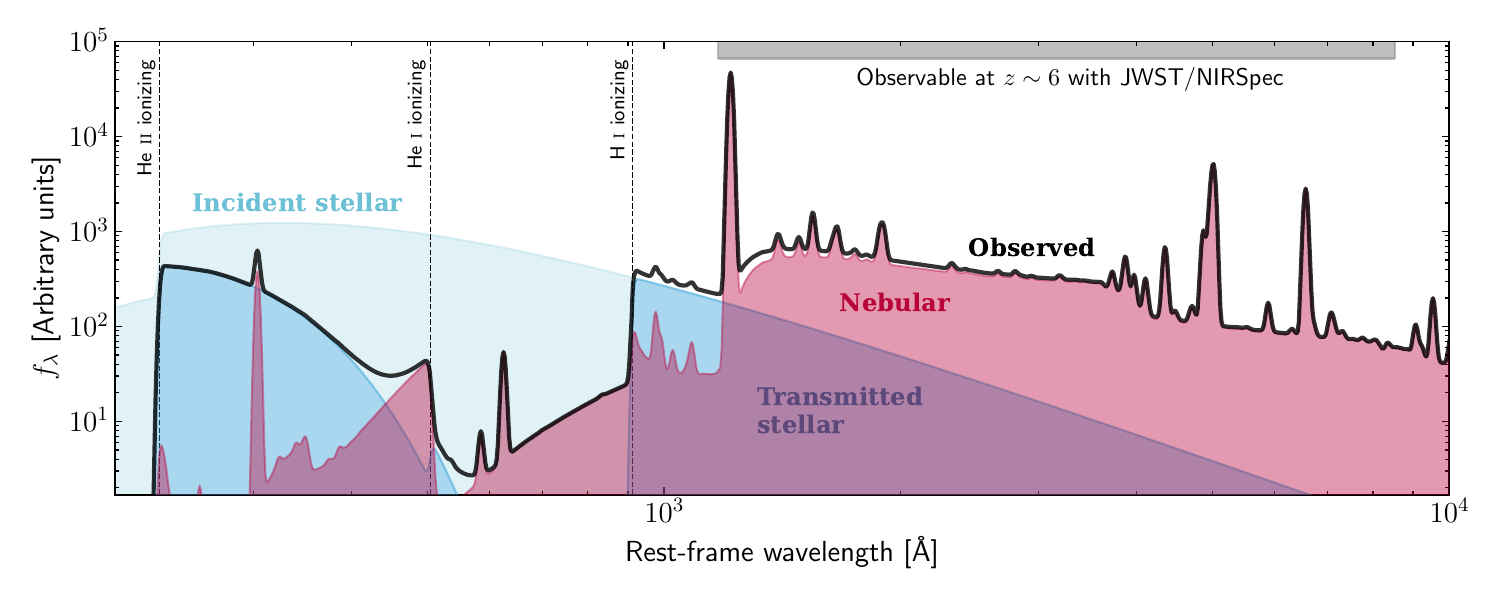}
    \caption{
    Schematic of how the hot star scenario gives rise to a nebular-dominated spectrum in a simple spherical nebula.
    The incident stellar spectrum ($T_{\rm eff}=10^{5.05}$ K) is shown in blue. The portion of this transmitted through the nebula is shown in the darker shade, while the lighter shade shows the portion that is absorbed to photoionisation of the gas. The red shows the resulting nebular spectrum, with the black showing the sum of nebular and transmitted stellar, which is what an observer will see.
    For a fixed 1500~\AA{} flux density, hot stars with $T_{\rm eff}\gtrsim10^{5}$ K emit significantly more ionising photons (light blue) than a typical stellar population, which in turn allows them to power much stronger nebular emission (red), which ultimately outshines their own spectrum at wavelengths longer than Ly$\alpha$ (1216~\AA{}).
    Note that a significant population of old stars can be accommodated within this scenario without affecting the nebular dominance at UV wavelengths, since these old stars primarly contribute flux at longer wavelengths.}
    \label{fig:visualisation_spectrum}
\end{figure*}

In summary, we identify three classes of model stellar SED that have the surface temperatures required to reproduce the observed two-photon continuum turnover in \objA\ (Figure~\ref{fig:best_fit_blackbody}; see also schematic in Figure~\ref{fig:visualisation_spectrum}). Wolf-Rayet stars with $Z=0.07~Z_\odot$, equal to the measured nebular metallicity, and hot massive stars with low-metallicity atmospheres provide a remarkably good fit, while existing model SEDs from stripped stars only fall short in having too strong suppression of He$^+$-ionizing flux. While a perfect match to the observed spectrum of \objA{} will require fine-tuning of the photoionization model, the fact that the continuum shape and the vast majority of the emission lines can be reproduced under very simple assumptions is encouraging that stars such as those considered in this section may be present in \objA.

\begin{table}
\caption{Parameters for the optimized blackbody model.}
\centering
\begin{tabular}{c|c}
    \hline
    Parameter & Value \\
    \hline
     n             & $10^3\ {\rm cm^{-3}}$ \\
     $T_{\rm BB}$  & $10^{5.05}$~K \\ 
     log $U$             & 0.0 \\
     $T_{\rm gas}$ & $10^{4.3}$~K \\
     $Z_{\rm O}/Z_{\odot}$ & $-1.1$ \\
     $Z_{\rm C}/Z_{\odot}$ & $-1.6$ \\
     O32           & 37.1 \\
     He~{\sc ii} leakage & 0.25 \\
     \hline
\end{tabular}
\label{tab:bb_params}
\end{table}


\section{Discussion} \label{sec:discussion}

\subsection{Implications for the stellar initial mass function if \objA{} hosts a population of hot stars}
\label{sub:imf}

We now consider what our modelling implies for the mass distribution of the stellar population in \objA.
The progenitor masses of Wolf-Rayet stars at the metallicity of \objA\ are not well constrained \citep{Massey2001}. Masses of $\geq37\ {\rm M_{\odot}}$ have been estimated for Wolf-Rayet stars in the Small Magellanic Cloud (SMC) with $T_{\rm eff}\gtrsim100,000$~K \citep{Hainich2015}, implying even higher progenitor masses. Meanwhile, the low-metallicity massive star models have somewhat higher progenitor masses of $\sim100~{\rm M_{\odot}}$.
While our search may not be exhaustive, we conclude from Figure~\ref{fig:spectral_overlay} that the nebular-dominated spectrum of \objA\ is consistent with ionization powered by low-metallicity massive stars ($\gtrsim 50~{\rm M_{\odot}}$), perhaps in the Wolf-Rayet phase.

If we adopt a progenitor mass of $50-100~{\rm M_{\odot}}$, then, assuming a typical IMF, we expect to form one such star per every $\sim1,300-3,800~{\rm M_{\odot}}$ of stellar mass.
To explore whether our model is consistent with this, we repeat our photoionization modelling with our best-fit massive star models while simultaneously adding a second component from a BPASS model.
We assume the BPASS stellar population has the same metallicity as the gas.
In the case of low-metallicity massive stars, we assume the BPASS model has an age of 3~Myr (the approximate lifetime of massive stars), while for the Wolf-Rayet models, we assume the progenitor stars are lower mass (50~M$_{\odot}$) and live for slightly longer (5~Myr).
We then progressively increase the ionization parameter of the second population until the UV turnover from the two-photon continuum begins to weaken. We calculate the stellar mass of a BPASS SSP that can be present per hot star as
\begin{equation}
    \label{eqn:imf}
    M_{\rm SSP}=\frac{U_{\rm SSP}}{U_{\rm star}}\frac{Q_{\rm star}}{Q_{\rm SSP}},
\end{equation}
where $Q_{\rm star}=10^{49.36}\ \gamma\ {\rm s^{-1}}$ for the Wolf-Rayet star model or $10^{49.98}\ \gamma\ {\rm s^{-1}}$ for the low-metallicity massive star model with $T_{\rm eff}=97,352$~K. 
$Q_{\rm SSP}=10^{46.73}$ or $10^{46.03}\ \gamma\ {\rm s^{-1}M_{\odot}^{-1}}$ at 3~Myr and 5~Myr, respectively. 
In both cases, we find the maximal allowable contribution to be $\frac{U_{\rm SSP}}{U_{\rm star}}\leq6.3\%$. Therefore, we can place upper limits of the BPASS SSP mass of $M_{\rm SSP}\leq112$ or 137~M$_{\odot}$ per one low-metallicity massive star or Wolf-Rayet star, respectively. 
This is clearly discrepeant from the values of $3,800~{\rm M_{\odot}}$ and $1,300~{\rm M_{\odot}}$ implied from a typical IMF.
In other words, there is a $35\times$ excess in the number of massive stars in the case of our low-metallicity massive star model, or a $9.5\times$ excess for our Wolf-Rayet model.
Either scenario implies that the IMF in \objA\ is very top-heavy.

We note that the measurement presented here is sensitive to the IMF by way of constraining the ratio between the ionising photon production rate, dominated in our model by stars more massive than $\gtrsim~50$~M$_{\odot}$, and the continuum flux at $\sim1500$~\AA{}, which in standard young stellar population models is dominated by stars with ${\rm M}\gtrsim5~{\rm M}_{\odot}$ \citep[e.g.][]{Byler2017}. It is, therefore, only sensitive to the high-mass end (${\rm M}\gtrsim5~{\rm M}_{\odot}$) of the IMF, and is completely insensitive to the impact of stars with ${\rm M}\lesssim5~{\rm M}_{\odot}$.

Taken at face value, this calculation implies that the ratio of $\gtrsim50$~M$_{\odot}$ to $\sim5-50$~M$_{\odot}$ stars in \objA\ is of order $\sim$10$\times$ higher than that predicted by a typical IMF.
However, the results of this calculation are very sensitive to the chosen underlying SSP, the assumed age, the progenitor masses of Wolf-Rayet stars, and the mass of the low-metallicity massive stars. For SSPs that have intrinsically lower $\xi_{\rm ion}$, the excess drops as $Q_{\rm SSP}$ appears in the denominator of Equation~\ref{eqn:imf}. If the ages at which the Wolf-Rayet stars evolve off the main sequence is longer than what we have assumed here, the excess will also decrease.
More generally, we highlight that the origin of such hot stars in high-redshift environments and the modelling of their atmospheres is highly uncertain and motivates detailed characterisation of the effects of binary stripping and stellar wind mass loss, particularly at low metallicity \citep{Gotberg2019, Vink2022}.
Thus, while we have demonstrated qualitatively that explaining the observed spectrum without variations to IMF is extremely difficult, a detailed characterisation of the high-mass end of the IMF in \objA\ will require more advanced understanding of stellar evolutionary processes in these environments.

Understanding the shape, upper-mass, and lower-mass cutoff of the IMF, and whether these evolve with initial conditions, is critical to the interpretation of nearly all extragalactic observables \citep{Hopkins2018}.
Theoretical works have predicted the IMF to get progressively more top-heavy for low-metallicity gas at high pressure \citep{Chon2021, Chon2022, Sneppen2022}, while others indicate that increased CMB temperature can cause the IMF to become more bottom-light at high-metallicity  \citep{Bate2023}.

While many studies of local field stars and young clusters have found no strong evidence for variation in the IMF of these systems (e.g. \citealt{Bastian2010} and references therein), a number of lines of evidence have emerged suggesting the IMF may vary in some environments.
Top-heavy IMFs have been derived in some local massive young star clusters \citep{Kalari2018, Schneider2018}, while some models of globular clusters have invoked a top-heavy IMF at early times to explain how gas expulsion proceeded from the young cluster \citep{Marks2012}. 

Spectral line indices in the spectra of early-type galaxies have been widely demonstrated to show that the IMF is bottom-heavy in these systems \citep{vanDokkum2010, Spiniello2014, MartinNavarro2015, Conroy2017, MM2024}. 
Similar conclusions have been drawn from gravitational lensing studies \citep{Treu2010, Smith2020} and dynamical modelling \citep{Cappellari2012, Poci2022, Lu2023}, both of which are senstive to the mass-to-light ratios.
We note that these measurements are sensitive to the ratios of stars with $M\lesssim0.5 M_{\odot}$ and $M\sim0.5-1 M_{\odot}$, which our constraint is insensitive to.
Furthermore, we note that chemical evolution modelling of massive ellipticals has shown that a time-invariant bottom-heavy IMF cannot explain the observed [Mg/Fe] and [Fe/H] abundance ratios in these systems with these studies instead suggesting that this bottom-heavy phase may have been preceeded by a short-lived top-heavy phase \citep{Vazdekis1997, Weidner2013, Jerabkova2018, DeMasi2019}.

Abundances of CNO elements and their isotopes are expected to be highly sensitive to variations in the high-mass end of the IMF \citep{Romano2022}.
Variations in the $^{13}$C/$^{18}$O ratio observed in $z\sim2-3$ dusty starburst galaxies have been interpreted as arising from a top-heavy IMF \citep{Zhang2018}, while a sub-solar C/O abundance ratios arising from a top-heavy IMF has been suggested to explain anomalous IR emission line ratios at high-redshift \citep{Katz2022_cii}.
Furthermore, a rapid starburst with a top-heavy IMF has been invoked as a possible explanation the enhanced N/O abundance ratio observed in GN-z11 \citep{Bekki2023}.

Top-heavy IMFs with low mass-to-light ratios have also been widely invoked as a possible cause of the surprising abundance of UV-bright galaxies observed at high redshift \citep[e.g.][]{Finkelstein2023, Harikane2024, Yung2024}.

\subsection{Total cluster mass and number of hot stars}

The hydrogen ionising photon luminosity, $Q$, can be approximated from the H$\beta$ flux as:
\begin{equation}
    Q \approx \frac{4\pi D_l^2 I_{\rm H\beta}}{(1-f_{\rm esc})h\nu_{\rm H\beta}}\frac{\alpha_{B}}{\alpha_{\rm B}^{\rm eff}},
\end{equation}
where 
$D_l$ is the luminosity distance, $I_{\rm H\beta}$ is the measured H$\beta$ flux, $h$ is Planck's constant, $\nu_{\rm H\beta}$ is the frequency of the H$\beta$ transition, $f_{\rm esc}$ is the escape fraction of ionising photons, $\alpha_{B}$ is the total case B recombination rate, and $\alpha_{\rm B}^{\rm eff}$ is the effective H$\beta$ recombination rate. To calculate a luminosity distance, we adopt a cosmology with $H_0=67.31$~km~s$^{-1}$~Mpc$^{-1}$ and $\Omega_{\rm m}=0.315$ \citep{Planck2016}. This gives $D_l=58.5$~Gpc, implying an H$\beta$ luminosity of $L_{\rm H\beta}=7.0\times10^{41}$ erg s$^{-1}$. 
Using the escape fraction derived from our best fitting blackbody photoionization model of $7.3\%$ and recombination rates evaluated at 20,000~K \citep{Osterbrock2006}, 
we estimate a hydrogen ionising photon luminosity of $Q=1.65\times10^{54}$ s$^{-1}$.

This value allows us to estimate the mass of the star clusters that have formed in \objA. 
The Wolf-Rayet star models presented in Section~\ref{sub:imf} have a fixed luminosity of $10^{5.3} L_\odot$, which results in a hydrogen ionising photon luminosity of $Q_{\rm WR}=10^{49.36}$ s$^{-1}$. However, known Wolf-Rayet stars in the most comparable local environments typically have significantly higher luminosities of $\sim10^{6.2} L_\odot$ \citep{Shenar2016}, implying that $Q_{\rm WR}\approx10^{50.3}$ s$^{-1}$ might be a more realistic value.
Based on the H$\beta$ luminosity, this would imply $\sim10,000$ of the $Z=0.07~Z_\odot$ Wolf-Rayet stars would be needed to power the spectrum of \objA.
In the alternative model, $\sim17,000$ very metal-poor stars of $\sim100$~M$_{\rm \odot}$ with $T_{\rm eff}=97,000$~K would be required.
Based on the calculations presented in Section~\ref{sub:imf}, this implies a maximum star cluster mass of $10^{6.22}\ {\rm M_{\odot}}$ for the Wolf-Rayet star model, or 
$10^{6.55}\ {\rm M_{\odot}}$ for the low-metallicity massive star model.
We caution that these mass estimates should be treated only as a guide since they are very sensitive to the adopted hot star SED as well as the same assumptions outlined in Section~\ref{sub:imf}.
We also note that the quoted value includes only the mass of recently formed population contributing significant UV luminosity. A considerable mass of older stellar populations could also be accommodated within these models.

\subsection{Are there other galaxies like \objA?}
\label{sub:other_objects}
While the physics driving the abnormal continuum shape of \objA{} remains uncertain, it is important to understand whether \objA{} is unique, or whether other objects show similar spectral features. Identifying larger samples of objects that are similar to \objA{} will help rule out scenarios that require fine-tuning. 

\begin{figure}
    \centering
    \includegraphics[width=0.47\textwidth]{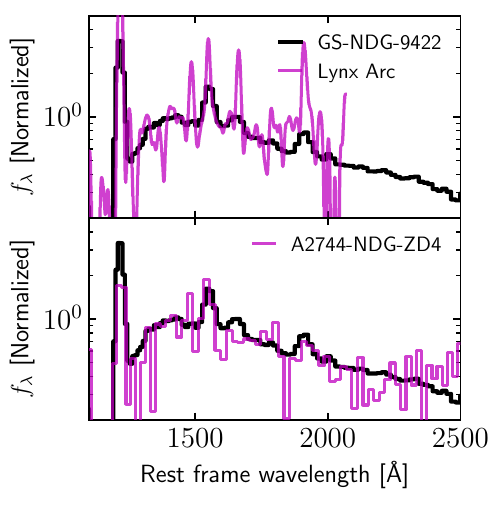}
    \caption{Zoom in on the rest-frame UV region for \objA\ (black) compared with the spectra of the $z=3.35$ Lynx arc (top) and \objB\ at $z=7.88$ (bottom). We have normalised the spectra and shifted them all to the rest frame.}
    \label{fig:uv_compare}
\end{figure}

While we have not performed an exhaustive or systematic search, in Figure~\ref{fig:uv_compare} we highlight two objects that both show Ly$\alpha$ and a UV turnover, similar to \objA{}. The first is the Lynx arc \citep{Fosbury2003}, a gravitationally-lensed, extreme emission system identified at $z=3.35$. The emission lines of the Lynx arc differ considerably from \objA{} in that He~{\sc ii} is weaker and there is strong N~{\sc iv}] emission which is not observed in \objA{}. Nevertheless the UV downturn and Ly$\alpha$ are remarkably similar. Indeed, analysis in \cite{Fosbury2003} came to similar conclusions as presented here for \objA\ -- namely, that hot stars ($T_{\rm eff}\gtrsim80,000$ K) are powering the emission in this galaxy which leads to the visibility of the two-photon continuum. Hence, it is clear that the shape of the continuum in \objA{} is not unique among the galaxy population.

We identify a second galaxy, \objB{} at $z=7.88$, observed as part of the Ultra-deep NIRCam and NIRSpec Observations Before the Epoch of Reionization (UNCOVER) program (ID: 2561, PI: Labbe; \citealt{Bezanson+22}) which also appears to show Ly$\alpha$ escape and a UV turnover (Figure~\ref{fig:uv_compare} bottom panel). The spectrum of this object, retrieved from  Dawn JWST Archive (DJA)\footnote{\url{https://dawn-cph.github.io/dja/index.html}} which had been reduced using the custom-made pipeline MsaExp v.0.6.12 \footnote{\url{https://zenodo.org/record/8319596}} (see \citealt{Heintz2023_DLA} for details), received only 2.3 hours integration. Thus, the continuum is not well detected in the rest-frame optical and we can not determine whether it shows a Balmer jump. Deeper data will be required to conclusively determine whether \objB{} is truly of the same nature as the Lynx arc and \objA.

\begin{figure*}
    \centering
    \includegraphics[width=0.9\textwidth]{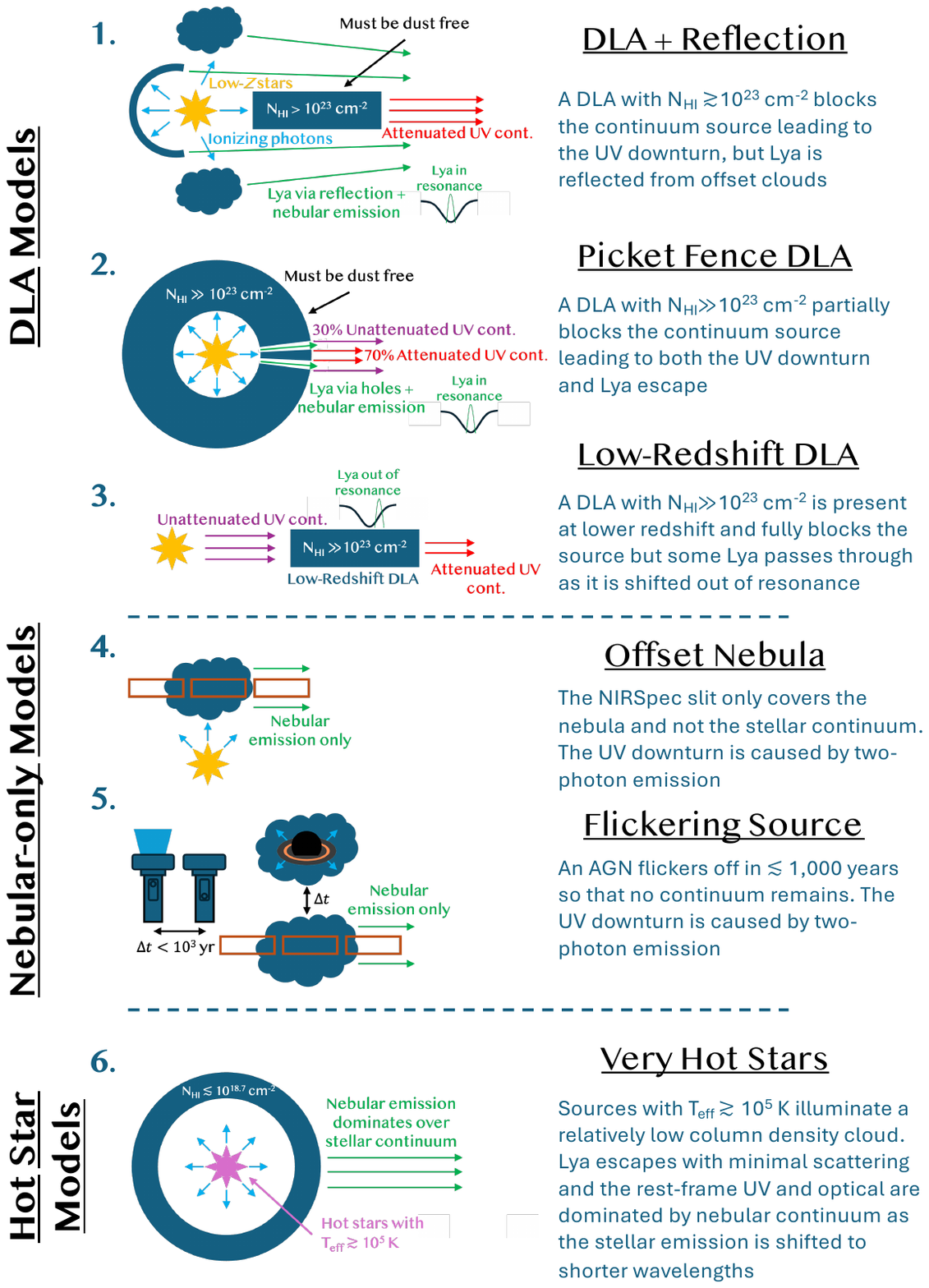}
    \caption{Summary of the three classes of scenarios that can explain the spectrum on \objA{}. For each scenario, we show various cartoons for how the physics may manifest and give rise to a steep UV slope, a high Ly$\alpha$ escape fraction, and the UV turnover redward of Ly$\alpha$. }
    \label{fig:cartoon}
\end{figure*}

\section{Conclusions} \label{sec:conclusion}

In this paper, we have presented a discussion of the possible physical origin of the observed continuum and line emission in \objA, a galaxy at $z=5.943$.
The identification of a spectral discontinuity of $15\pm0.9$ nJy at the location of the Balmer limit is clearly indicative of very strong nebular emission, as are the high equivalent widths observed for \OIIIl5007, H$\beta$, and H$\alpha$.
We measure high nebular temperatures ($(1.83\pm0.15)\times10^4$ K), low densities ($n_e\lesssim10^3$ cm$^{-3}$), and low metallicity ($12+\log({\rm O}/{\rm H})=7.59\pm0.01$; $\sim$8~\% (O/H)$_\odot$) in this system.
We show that, considering only the emission line ratios, this system is highly consistent with emission powered by a typical population of young, metal-poor stars, with no strong evidence for the presence of AGN activity.

What is more difficult to explain is the observed shape of the continuum. In particular, the strong UV turnover observed at $\lambda_{\rm rest}\approx1430$~\AA{}.
The shape of this feature is a remarkably consistent with the hydrogen two-photon continuum, but could also be explained by a very high column density of neutral hydrogen. We consider several scenarios for how each of these effects could self-consistently give rise to the observed spectrum of \objA, all of which are summarized in Figure~\ref{fig:cartoon}.

The primary challenge faced by models explaining this feature with a DLA is that the magnitude of the turnover implies unprecedented column densities of $N_{\rm HI}\gtrsim10^{23}$ cm$^{-2}$, higher than any known DLA. 
Furthermore, the observation of narrow, high EW Ly$\alpha$ emission with a high escape fraction ($\sim27\%$) requires some mechanism by which this emission can reach the observer.
This can be reconciled if Ly$\alpha$ arises from a reflection spectrum (Scenario \#1), although known examples of this are orders of magnitude larger in physical scale than \objA.
Alternatively, the Ly$\alpha$ might be allowed to reach the observer if the DLA has only partial coverage (Scenario \#2) or is at lower redshift (Scenario \#3); however, each of these scenarios imply even higher neutral gas column densities which means their plausibility is questionable. Why such high gas columns would not be fully molecular and why at this moderate metallicity there are no signatures of dust attenuation with so much neutral gas remain open questions in the DLA scenario. For these reasons, we do not favour the DLA solution.

We demonstrated in Figure~\ref{fig:neb_cont_fitting} that modelling the continuum of \objA\ with strong two-photon, free-bound and free-free continuum plus a subdominant young stellar and old stellar component provides a very good fit to the spectrum, with the predicted nebular temperature and equivalent widths of Balmer emission highly consistent with what is measured from the emission lines. The main challenge for explaining the UV turnover with two-photon continuum emission is therefore simply: how can the system power such strong nebular emission without the rest-UV continuum becoming dominated by the spectrum of the ionising source?

One possibility is that no flux from the ionising source is observed, either due to a spatial offset whereby the ionising source falls outside the slit (Scenario \#4) or because the ionising source has recently turned off and we are observing emission from its relic (Scenario \#5).
The former is disfavoured by the compact morphology of \objA, which is well-centred in the slit (Figure~\ref{fig:spectrum}).
The predicted recombination timescales in Scenario \#5 ($\sim$500 - 5,000 yr for H$^+$ and $\sim$30-300 yr for He$^{++}$) are incredibly short meaning we would need to have caught this object at a very specific moment in time, although this could perhaps be explained with an AGN fading on a timescale of $\lesssim10^4$ yr \citep[e.g.][]{French2023}. We consider this model to be possible, but the identification of larger samples of similar may rule this scenario out.

The final scenario we consider is that the ionising source is indeed present in the slit, but its ionising photon production efficiency is so large that it powers sufficiently strong nebular emission that the two-photon continuum provides the dominant contribution to the rest-UV continuum (Scenario \#6 and Figure~\ref{fig:visualisation_spectrum}).
This is a generic prediction of all photoionization models that include a significant population of hot stars with $T_{\rm eff}\gtrsim10^5$~K \citep[e.g.][]{Schaerer2002,Raiter2010,Zackrisson2011,Inoue2011,Trussler2023}. We identify several model stellar SEDs that can attain these temperatures, showing that models with ionising spectra dominated by $\sim$7~\% $Z_\odot$ Wolf-Rayet stars and/or very low-metallicity massive stars are able to reproduce the puculiar shape of the observed continuum.
The critical aspect to this scenario is that it requires a top-heavy IMF, with the implied ratio of $\gtrsim$50~M$_\odot$ to $\sim$5-50~M$_\odot$ stars significantly enhanced relative to typical stellar populations.
We note that the precise magnitude of this excess is highly dependent on the assumptions around the evolution and atmospheric properties of these hot, massive, low-metallicity stars, and this work motivates a pursuing improved models of these stars.
Nonetheless, we note that although such an IMF is unprecedented in nearby star clusters, it is broadly consistent with other suggestions of enhanced massive star formation in extreme systems at early times (see discussion in Section~\ref{sub:imf}).

All three classes of solutions we have proposed to explain the spectrum of \objA\ pose interesting physical questions about high-redshift galaxy formation. Either we have discovered the highest DLA column density for a star-forming galaxy, we have witnessed the flickering of a powerful ionising source on timescales much shorter than typical QSO lifetimes, or the ionizing source represents an exotic population of stars with high surface temperature. The similarities with the Lynx Arc at $z=3.35$ and \objB{} at $z=7.88$, indicate that \objA{} may not be unique. This motivates a more systematic search of such objects to unravel the origin of their unique spectra.

\section*{Acknowledgements}

We thank Bob Fosbury for useful discussions and providing us with the LRIS spectrum of the Lynx arc. We thank Paul Crowther for providing insightful and constructive comments.
We thank Thibault Garel for providing us with the simulated IGM attenuated curves from the SPHINX simulation. We thank Anna Feltre and St\'ephane Charlot for sharing their latest AGN photoionisation model grids. AJC, AS and AJB, acknowledge funding from the ``FirstGalaxies'' Advanced Grant from the European Research Council (ERC) under the European Union's Horizon 2020 research and innovation programme (Grant agreement No. 789056). CW thanks the Science and Technology Facilities Council (STFC) for a PhD studentship, funded by UKRI grant 2602262. NL acknowledges support from the Kavli foundation. 

\section*{Data Availability}

Reduced spectra, imaging, and measured photometry of \objA\ were made publicly available as part of JADES Data Release 1 \citep{Bunker2023_DR, Rieke2023} and can be found at \url{https://archive.stsci.edu/hlsp/jades}. 
Photoionisation models presented in this work will be shared on reasonable request.



\bibliographystyle{mnras}
\bibliography{paper} 




\appendix

\section{Empirical DLA Column Density Measurements}
\label{app:dla_fitting}

In Section~\ref{sec:physical_conditions}, we optimized a  {\small CLOUDY} photoionization model in order to reproduce the emission lines of \objA, which then provided an estimate of the continuum. We then applied various DLA columns with leakage to this model in order to reproduce the UV turnover and the escape of Ly$\alpha$. However, the exact column density needed is degenerate with the underlying shape of the continuum. In this section, we repeat the experiment trying only to match the shape of the continuum.

Following the approach in Section~\ref{sec:continuum}, we run an MCMC to fit the continuum of \objA. In this case, we do not allow the two-photon continuum to have a free normalization (and hence we remove this parameter), and we replace it with a variable DLA column density and leakage. The model thus has eight free parameters (compared to the seven used in Section~\ref{sec:continuum}).

\begin{figure}
    \centering
    \includegraphics[width=0.45\textwidth]{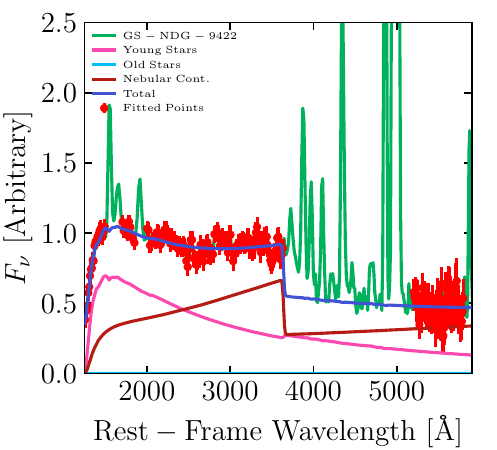}
    \includegraphics[width=0.45\textwidth]{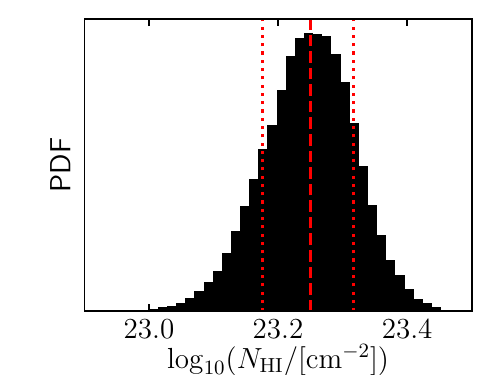}
    \includegraphics[width=0.45\textwidth]{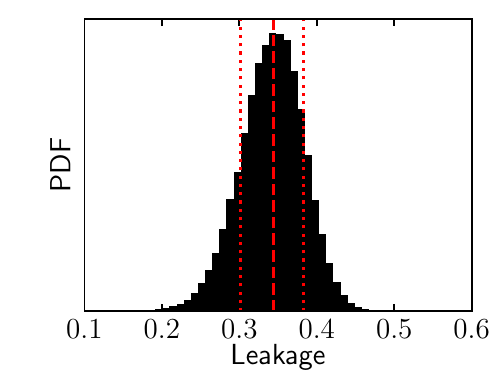}
    \caption{
    {\it Top:} Best-fit continuum model from fitting described in Appendix~\ref{app:dla_fitting}. Young stellar component + DLA is shown in pink, nebular component with a fixed low two-photon continuum contribution is shown in brown, while the old stellar component is shown in light blue.
    {\it Middle:} Resulting posterior distribution of neutral hydrogen column density.
    {\it Bottom:} Resulting posterior distribution of leakage fraction.
    }
    \label{fig:app_dla}
\end{figure}

In Figure~\ref{fig:app_dla} we show the fit corresponding to the 50th percentile model in the posterior distributions of each parameter as well as the marginalized posterior distributions on H~{\sc I} column density and leakage. Indeed we find values close to $10^{23.3}\ {\rm cm^{-2}}$, which is slightly lower than reported in Section~\ref{sec:physical_conditions}, but very consistent within $1\sigma$. In either case, the DLA column required is exceptionally high compared to known DLAs.

Interestingly, despite the fact that Ly$\alpha$ was not included in the fitting, the MCMC tightly constrains the amount of leakage to be $\sim35\%$. This is because the shape of the transmission curve for the DLA is inconsistent with the shape of the UV downturn in \objA, while in contrast, it is much more consistent with that of the two-photon continuum. Nevertheless, it is encouraging to see that the MCMC requires the same leakage as discussed in Section~\ref{sec:physical_conditions}.

While this DLA model is able to reproduce the shape of the continuum, we find two inconsistencies. First, the 50th percentile temperature of 22,369~K is much higher than that measured by the [O~{\sc iii}]~$\lambda$4363 auroral line. As stated above, these two temperatures do not necessarily have to agree, but are typically observed to be close \citep{Guseva2006,Guseva2007}. Moreover, in this model, the predicted H$\beta$ EW is 767~\r{A}$\pm40$~\r{A} which is very inconsistent with what is observed. This is in stark contrast to the model where the UV is dominated by two-photon emission. Hence the results in this section confirm and strengthen our conclusions that the spectrum of \objA{} is more consistent with being primarily nebular in origin.

\section{Models with XRBs}
\label{app:xrbs}

\begin{figure}
    \centering
    \includegraphics[width=0.45\textwidth]{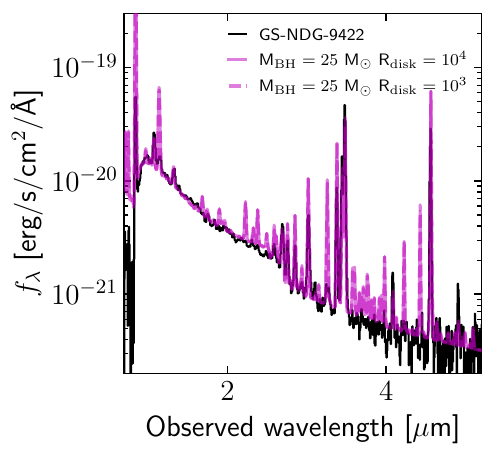}
    \caption{Spectrum of \objA\ (black) compared with photoionization models that include high-mass X-ray binaries with a black hole mass of $25\ {\rm M_{\odot}}$ (magenta). The solid and dashed magenta lines indicate models with different accretion disk radii.}
    \label{fig:xrb}
\end{figure}

During the course of the modelling presented in Section~\ref{sec:spectral_modelling}, we also considered the possibility that the emission is powered by X-ray binaries.
Following \citet{Senchyna2020} and \citet{Katz2023}, we model the X-ray sources using a modified blackbody spectrum \citep{Mitsuda1984}. We assume black hole masses in the range $6-25\ {\rm M_{\odot}}$ and disk radii between $10^3-10^4$~gravitational radii. In order for a UV turnover to appear, the ionizing photon output from the XRBs must dominate over the stellar population so that the nebular continuum can outshine the stars, implying a very high ratio of X-ray luminosity to star formation rate. In Figure~\ref{fig:xrb}, we show our XRB model for 25~M$_{\odot}$ black holes optimised to reproduce the continuum shape of \objA. This model significantly over-predicts the strength of the He~{\sc ii}~$\lambda$1640 and $\lambda$4686 lines. 
Varying the black hole masses does not resolve this issue.
We therefore conclude that XRBs are not the dominant ionizing source in \objA.

\bsp	
\label{lastpage}
\end{document}